\RequirePackage{lineno}
\documentclass[aps,prd,superscriptaddress,groupedaddress]{revtex4}
\usepackage{graphicx}
\usepackage{dcolumn}
\usepackage{bm}
\usepackage{amssymb}
\usepackage{multirow}
\usepackage{epsfig}
\usepackage{amsmath}
\usepackage{float}
\usepackage{color}
\usepackage[colorlinks, linkcolor=blue]{hyperref}
\usepackage{lineno}

\newcommand{\effstw}{\ensuremath{\sin^2\theta_{\text{eff}}^{\text{$\ell$}}}}

\newcommand{\stw}{\ensuremath{\sin^2 \theta_{W}}}

\newcommand{\eP}{\texttt{ePump }}

\begin{document}

%\linenumbers
\lefthyphenmin=2
\righthyphenmin=2

\widetext

\title{Reduction of PDF uncertainty in the measurement of the weak mixing angle at the ATLAS experiment}
\affiliation{Department of Modern Physics, University of Science and Technology of China, Jinzhai Road 96, Hefei, Anhui, 230026, China}
\author{Yao Fu}\affiliation{Department of Modern Physics, University of Science and Technology of China, Jinzhai Road 96, Hefei, Anhui, 230026, China}
\author{Siqi Yang‰}\affiliation{Department of Modern Physics, University of Science and Technology of China, Jinzhai Road 96, Hefei, Anhui, 230026, China}
\author{Minghui Liu‰}\affiliation{Department of Modern Physics, University of Science and Technology of China, Jinzhai Road 96, Hefei, Anhui, 230026, China}
\author{Liang Han‰}\affiliation{Department of Modern Physics, University of Science and Technology of China, Jinzhai Road 96, Hefei, Anhui, 230026, China} 
\author{Tie-Jiun Hou}\affiliation{Department of Physics, College of Sciences, Northeastern University, Shenyang 110819, China}
\author{Carl Schmidt}\affiliation{Department of Physics and Astronomy, Michigan State University, East Lansing, MI 48823, USA}
\author{Chen Wang}\affiliation{Department of Modern Physics, University of Science and Technology of China, Jinzhai Road 96, Hefei, Anhui, 230026, China}
\author{C.--P. Yuan}\affiliation{Department of Physics and Astronomy, Michigan State University, East Lansing, MI 48823, USA}

\begin{abstract}

We investigate the parton distribution function (PDF) uncertainty in the measurement of the effective weak mixing angle $\effstw$ at the CERN 
Large Hadron Collider (LHC). 
The PDF-induced uncertainty is large in the proton-proton collisions at the LHC due to the dilution effect. 
The measurement of the Drell-Yan forward-backward asymmetry ($A_{FB}$) at the LHC can be used to reduce the PDF uncertainty in the $\effstw$ measurement. However, when including the full mass range of lepton pairs in the $A_{FB}$ data analysis, the correlation between the PDF updating procedure and the $\effstw$ extraction leads to a sizable bias in the obtained $\effstw$ value. 
From our studies, we find that the bias can be significantly reduced by removing Drell-Yan events with invariant mass around the $Z$ pole region, while most of the sensitivity in reducing the PDF uncertainty remains. 
Furthermore, the lepton charge asymmetry in the $W$ boson events as a function of the rapidity of the charged leptons, $A_\pm(\eta_\ell)$, is 
known to be another observable which can be used to reduce the PDF uncertainty in the $\effstw$ measurement. 
The constraint from $A_\pm(\eta_\ell)$ is complementary to that from the $A_{FB}$, thus no bias affects the $\effstw$ extraction. 
The studies are performed using the Error PDF Updating Method Package ({\sc ePump}), which is based on the Hessian updating methods. In this article, the CT14HERA2 PDF set is used as an example. 

\end{abstract}

\maketitle

\section{Introduction}

Measurement of the leptonic effective weak mixing angle, $\theta^\ell_\text{eff}$, is one of the most important topics in experimental particle physics. It is the key parameter 
in electroweak global fitting. It played a crucial role in predicting the mass of the Higgs boson with a precision of {\cal O}(10) GeV. Going forward, it will continue to contribute in the global fittings, and will aid in tests of the standard model and in searches for potential new physics beyond the standard model.
At an energy scale of the $Z$ boson mass ($M_Z$), $\effstw$ can be determined from measurements of parity-violation in the neutral-current 
processes of fermion-antifermion scattering, $f_i\bar{f_i}\rightarrow Z/\gamma^* \rightarrow f_j\bar{f_j}$.  One such measurement is the forward-backward asymmetry 
($A_{FB}$), defined as
\begin{eqnarray}
A_{FB} = \frac{N_F - N_B}{N_F + N_B}\,,
\end{eqnarray}
where $N_F$ and $N_B$ are the numbers of forward and backward events.
At lepton colliders, forward and backward events are defined according to the sign of  $\cos\theta$, where $\theta$ is the scattering angle between 
the outgoing fermion $f_j$ and the incoming fermion $f_i$.
The most precise determinations to date of $\effstw$ at the $Z$-pole are provided by the LEP and SLD Collaborations~\cite{LEP-SLD}, 
giving a combined result of $0.23153 \pm 0.00016$. The precisions of these measurements, achieved at the last generation of $e^+e^-$ colliders, 
are limited by statistical uncertainties. 
Subsequent to the LEP/SLD era, measurements have been made at hadron collider experiments, {\it i.e.,} the proton-antiproton collider, Tevatron, and the proton-proton collider, CERN Large Hadron Collider (LHC), using 
$A_{FB}$ in the final states of Drell-Yan (DY) $p\bar{p}/pp\rightarrow Z/\gamma^* \rightarrow \ell^+\ell^-$ processes, as a function 
of the di-lepton pair invariant mass. 
At hadron colliders, forward and backward events are defined in the Collins-Soper (CS)~\cite{CS-frame}. This is a special rest frame of the lepton-pair, with the polar and azimuthal angles defined relative to the two hadron beam directions. The $z$ axis is defined in the $Z$ boson rest frame so that it bisects the angle formed by the momentum of either of the incoming hadron and the negative of the momentum of the other hadron. 
The cosine of the polar angle $\theta^{*}$ is defined by the direction of the outgoing lepton $l^{-}$ relative to the $\hat{z}$ axis in the CS frame and can be calculated directly from the laboratory frame lepton quantities by

\begin{eqnarray}
\cos\theta_{CS}^{*} = c \, \frac{2(p^{+}_{1}p^{-}_{2} -
	 p^{-}_{1}p^{+}_{2})}
{m_{ll}\sqrt{m^2_{ll} + p^{2}_{{\rm T},ll}}} \,,
\end{eqnarray}
where the scalar factor $c$ (either 1 or -1) is defined for the Tevatron and the LHC, respectively, as 
\begin{eqnarray}
c =\Bigl\{
\begin{array}{ll}
1, &   \mbox{for the Tevatron} \\
\vec{p}_{Z,ll}/|\vec{p}_{Z,ll}|, & \mbox{for the LHC}\,. \\
\end{array}
\label{eq:zdirection}
\end{eqnarray}
And thus, the sign of the $z$ axis is defined as the proton beam direction for the Tevatron, and on an event-by-event basis as the sign of the lepton pair 
momentum with respect to the $z$ axis in the laboratory frame for the LHC.
The variables $p_{Z,ll}$, $m_{ll}$, and $p_{{\rm T},ll}$ denote the longitudinal momentum, invariant mass and transverse momentum of the dilepton system, 
respectively, and,
\begin{eqnarray}
p^{\pm}_i = \frac{1}{\sqrt{2}}(E_i \pm p_{Z,i}) \,,
\end{eqnarray}
where the lepton (anti-lepton) energy and longitudinal momentum are $E_{1}$ and $p_{Z,1}$ ($E_{2}$ and $p_{Z,2}$), respectively. 
The DY events are therefore defined as forward ($\cos\theta_{CS}^{*}>0$) or backward ($\cos\theta_{CS}^{*}<0$) according to the direction of 
the outgoing lepton in this frame of reference.
 
Compared to the lepton-collider cases, measurements at hadron colliders suffer from additional uncertainties on modeling the directions of the incoming fermions 
and antifermions in the initial state. Such uncertainties will dilute $A_{FB}$ and reduce the sensitivity for the determination of $\effstw$. The degree of the dilution at hadron colliders is modeled by parton 
distribution functions (PDFs). At the Tevatron, fermions in the initial state of DY 
production are dominated by valence quarks. This allows us to make an assumption that the incoming quark of DY production is moving  
along the proton beam direction, as indicated in Eq.~\ref{eq:zdirection}, while the direction of the incoming anti-quark is along the anti-proton beam. However, contribution from sea-quark interactions 
is still as large as about $10\%$ at the Tevatron.  The uncertainty of this dilution fraction, which is calculated using PDFs, will propagate into the uncertainty estimation of the $\effstw$ measurement extracted 
form $A_{FB}$ distribution. The combination of the D0 and CDF measurements at the Tevatron gives a result of
$0.23179 \pm 0.00030(\text{stat})\pm 0.00017(\text{PDF}) \pm 0.00006 (\text{syst})$~\cite{Tevatron-combine}, 
which shows a non-negligible PDF-induced uncertainty.

The PDF dilution effect is even more significant at the LHC, since it is a proton-proton collider.
Due to its completely symmetrical initial state, there is an equal probability of finding the incoming quark of DY  production from either of the two proton beams. 
In order to distinguish forward from backward events in $pp$ collisions, 
the beam pointing to the same hemisphere as the $Z$ boson reconstructed from final state leptons, is assumed to be the one which provides the quark. 
This is motivated by the observation that the valence quarks inside the protons generally carry more energy than the antiquarks (or sea quarks) inside the protons. 
However, this assignment is only statistically correct, because it is possible for the sea quarks to have 
a larger fraction of momentum ($x$) of the incoming proton than the valence quarks. 
Furthermore, beyond the leading order in the QCD interaction, quark-gluon and antiquark-gluon processes will contribute at the next-to-leading order (NLO), and the gluon-gluon process will contribute at the next-to-NLO (NNLO). 
These all affect the PDF dilution factor, whose magnitude depends on the precise modeling of the momentum spectra of all flavors 
of quarks and gluons involved in the Drell-Yan processes, which is more complicated than just modeling the total cross sections of valence quarks and sea quarks for the proton-antiproton case.
Consequently, the PDF-induced uncertainty in the $A_{FB}$  measurement at the LHC 
is significantly larger than that at the Tevatron. The latest published measurement from the CMS collaboration 
gives a result of $0.23101\pm 0.00036(\text{stat}) \pm 0.00031(\text{PDF}) \pm 0.00024(\text{syst})$~\cite{CMS-8TeV}, in which the PDF uncertainty is about the same size as the statistical uncertainty. 

In the future high luminosity (HL) LHC era, the statistical uncertainty will be reduced as data accumulates. Thus, the PDF 
uncertainty will become the leading uncertainty that limits the precision in the determination of $\effstw$. 
Studies have been done in the literature to discuss how to further reduce the PDF uncertainties relevant for 
precision electroweak measurements at the LHC~\cite{ArieMassConstrain}.
Two experimental observables are essential to this task: one is the $A_{FB}$ of the DY pairs and the other is 
the lepton charge asymmetry $ A_\pm(\eta_\ell)$ in the $W^\pm$ boson events. 
When $A_{FB}$ is used to simultaneously determine $\effstw$ and to reduce PDF uncertainties, it will 
inevitably bring correlations. Such correlations have not been systematically considered in previous studies, as it is not expected to be important when the PDF-induced uncertainty does not dominate the overall uncertainty.
In this article, we investigate the correlation between the two tasks of further reducing the PDF uncertainty and performing the precision determination of $\effstw$ from measuring the same experimental observable $A_{FB}$.  
We demonstrate 
the potential bias on the $\effstw$ determination, and discuss possible solutions for 
the future LHC measurements. 

The paper is organized as follows: in Section II, a brief review on using new data to update PDFs and to reduce the related uncertainties  is presented; in Section III, we perform an exercise in updating the PDFs with $A_{FB}$ at the LHC, and demonstrate its potential bias on the $\effstw$ determination;
in Section IV, we study how updating the PDFs with 
the lepton charge asymmetry $ A_\pm(\eta_\ell)$ measured at the LHC reduce the PDF uncertainties. 
In Section V, we study the implications of updating the PDFs with both $A_{FB}$ and $ A_\pm(\eta_\ell)$ data, and 
apply the ePump-optimization procedure to illustrate the complimentary roles of 
the sideband $A_{FB}$ and $ A_\pm(\eta_\ell)$ observables in reducing the PDF uncertainty, and then to make the optimal choice on the bin size of the experimental data used in the PDF-updating analysis;   
finally, a summary is presented in Section VI.

\section{PDF updating method and $A_{FB}$}
\begin{figure}[hbt]
\begin{center}
\includegraphics[width=0.4\textwidth]{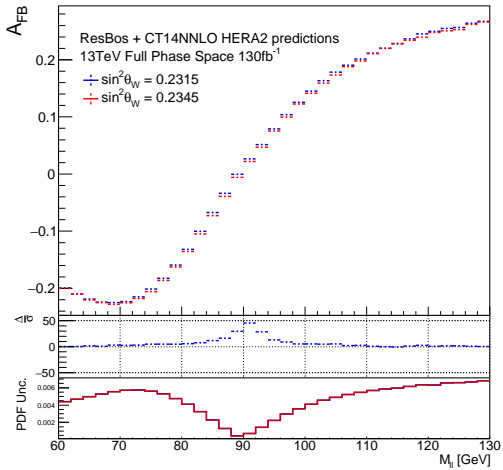}
\caption{\small Theory prediction of $A_{FB}$ as a function of $M_{\ell\ell}$ at 13 TeV LHC. 
The $\Delta/\sigma$ in the middle panel is defined as $(A_{FB}[\effstw=0.2345] - A_{FB}[\effstw=0.2315]) / \sigma$, 
where $\sigma$ is the statistical uncertainty of the event samples in that mass bin. The bottom panel shows the magnitude of PDF-induced uncertainty of $A_{FB}$, predicted by the CT14HERA2 error PDFs, at the 68\% CL.
}
\label{fig:AFBfullmass_comparestw}
\end{center}
\end{figure}

The two most commonly-used methods for extracting PDFs and their uncertainties from a global analysis of high-energy scattering data,  
are the Monte Carlo method, used by NNPDF~\cite{NNPDF3.0}, and the Hessian method, used in CT14HERA2~\cite{Dulat:2015mca,Hou:2016nqm}, for example.
In the Monte Carlo method,
a statistical ensemble of PDF sets are provided, which are assumed to approximate the probability distribution of possible PDFs, as constrained from the global analysis of the data.
In the Hessian method, a smaller number of error PDF sets are provided along with the central set which minimizes the $\chi^2$-function in a global analysis. These error PDF sets correspond to the positive and negative eigenvector directions in the space of PDF parameters.
The most complete method for obtaining constraints from the new data on the PDFs would be to add the new data into the global analysis package and to do a full re-analysis.  However, this is impractical 
for most users of the PDFs.
A technique for estimating the impact of new data on the PDFs, without performing a full global analysis, is very useful. 
In the context of the Monte Carlo PDFs, the PDF reweighting method has become commonplace.  This involves applying a weight factor to each of the PDFs in the ensemble~\cite{Giele:1998gw,Ball:2010gb,Ball:2011gg} when performing ensemble averages. 
The PDF updating procedure will reduce the overall effective number of PDF replica in the ensemble. 
The impact of new data can also be estimated directly using Hessian PDFs~\cite{Paukkunen:2014zia,Camarda:2015zba,ePumpPaper}, where it is called Hessian profiling.
It updates the eigenvectors within the Hessian approximation, which is faster and simpler. 
Note that both the Monte Carlo method and the Hessian profiling are based on the original Monte Carlo PDFs or error sets, respectively. 
Therefore, the new data is assumed to be in general consistent with the PDF 
predictions before updating, so that the updated best-fit PDF set is not too different from the original best fit. If a large deviation is found 
between the new data and the original theory predictions, a full analysis of PDF global fitting is needed. 

The theoretical predictions in this work are computed using the {\sc ResBos}~\cite{RESBOS} package at the next-to-leading order (NLO) plus the next-to-next-leading log (NNLL) in QCD, in which the canonical scales are used~\cite{Landry:2002ix,Su:2014wpa}.
The CT14HERA2 central and error PDFs~\cite{Dulat:2015mca,Hou:2016nqm} are used in this analysis. 
$A_{FB}$ as a function of dilepton mass ($M_{\ell\ell}$) at LHC is sensitive both to $\effstw$ and to PDF modeling. 
FIG.~\ref{fig:AFBfullmass_comparestw} shows the $A_{FB}$ distributions of two separated 
$\effstw$ values of 0.2315 and 0.2345, their difference, 
and the PDF uncertainties as functions of the di-lepton invariant mass for $\sqrt{s}=13$~TeV $pp$ collisions at the LHC.
The two values of $\effstw$ are 
arbitrarily chosen to be far separated in order to clearly reveal their different predictions of $A_{FB}$. 
When $A_{FB}$ from a new data set is used in the PDF updating procedure, it is assumed to be consistent with 
the current theory predictions. This means that $\effstw$, on which $A_{FB}$ depends, 
is considered to have the same value as determined from existing experimental measurements, 
even if a different value of $\effstw$ is used in generating the pseudo-data.
As a result, a simple PDF updating procedure will forcibly absorb the difference in 
$\effstw$ into the PDFs, which will bias the determination of both the updated PDFs and the extracted $\effstw$. 
The size of the bias depends on how large is the difference between the current accepted value of $\effstw$ (used in the theory prediction) and 
the value used in the generation of the pseudo-data, which will be quantitatively discussed in the following sections. An important thing to note is that $A_{FB}$ 
is more sensitive to $\effstw$ in the $Z$ pole region, while the PDF-induced uncertainty becomes more significant when $M_{\ell\ell}$ moves to higher or lower regions. 
The difference in sensitivities of the regions suggests a method to reduce the correlations.
The work presented in Ref.~\cite{ArieMassConstrain} was done using the  
Monte Carlo reweighting method, with NNPDF PDFs, and was based on the hypothesis that the above-mentioned correlation is negligible. 
In this work, we instead use the software package {\tt ePump} (error PDF Updating Method
Package), which can update any given set of Hessian PDFs obtained from an earlier global analysis~\cite{Willis:2018yln}. 

\section{Updating the PDFs with $A_{FB}$ data }

In this section, we quantitatively examine how the PDF-induced uncertainty in the determination of $\effstw$ could be reduced by applying the Hessian updating method, via {\tt ePump}, and study the 
correlation mentioned above. 
First, we consider the case of using the $A_{FB}$ data spanning the full  range of $M_{\ell\ell}$, from 60 GeV to 130 GeV. 
Second, we consider the case of using only the $A_{FB}$ sideband spectrum, where the events with $M_{\ell\ell}$ from 80 GeV to 100 GeV are excluded.

In order to perform the PDF update, \texttt{ePump} requires 
two sets of inputs: data templates and theory templates. The data templates provide $A_{FB}$ distributions with 
their uncertainties.
The theory templates consist of the theory predictions for the $A_{FB}$ from the original PDF error sets. 
The output of \texttt{ePump} consists of an updated central and Hessian eigenvector PDFs, representing 
the result that would be obtained from a full global re-analysis that includes the new data.  
As an additional benefit, \texttt{ePump} can also output the updated predictions and uncertainties for any other observables 
of interest without the necessity to recalculate using the updated PDFs.
For more details about the use of \texttt{ePump}, see Ref.~\cite{ePumpPaper}.

For the DY samples, 
each lepton flavor channel of electron and muon has 250 million events in the mass range of $60$ GeV $\le M_{\ell\ell} \le 130$ GeV. 
This sample size corresponds to an integrated luminosity of roughly 130 fb$^{-1}$, which is the size of the total data collected by the ATLAS 
detector during the LHC Run 2.
The pseudo-data 
is modeled using the \texttt{CT14HERA2} central PDFs. 
Nominal theory-template samples consist of the central and (56) error PDF predictions, generated using the \texttt{CT14HERA2} error PDFs sets. 
In the theory templates, $\effstw$ is set to be 0.2315, which is the value determined by the LEP and 
SLD Collaborations.  In the pseudo-data, $\effstw$ is set to be 0.2324 in order to examine the effects 
of an offset or pull in the new data.
The difference is deliberately chosen to be 3 times the uncertainty of the $\effstw$ measurement as determined at hadron colliders~\cite{Tevatron-combine}. 
To mimic the experimental acceptance, a set of ATLAS detector-like event selections are further applied to the pseudo-data and the nominal theory samples:
\begin{itemize}
	\item Each lepton is required to have its transverse momentum $p_T$ $\ge$ 25 GeV.
	\item Lepton pseudo-rapidity is limited by $|\eta_\ell|$ $<$ 4.9. 
	\item Events are denoted as $CC$ (central-central) if both leptons have  $|\eta_\ell| \le 2.5$, and $CF$ (central-forward) if one lepton 
	has $|\eta_\ell| \le 2.5$ and the other has $2.5<|\eta_\ell|<4.9$. The $CC$ events correspond to doubling the integrated 
	luminosity with respect to the $CF$ events, since both the dielectron and dimuon channels contribute to the $CC$ events,  
	while only the dielectron channel has $CF$ events at the ATLAS detector.
	\item The $Z$ pole region is defined as dilepton invariant mass satisfying  $80 \le M_{\ell\ell} \le 100$ GeV. The sideband region 
	is defined as $60 < M_{\ell\ell} < 80$ GeV and $100<M_{\ell\ell}<130$ GeV.
	\item The forward-backward asymmetry $A_{FB}$ is measured in a 2 GeV mass bin size. 
\end{itemize}
Note that the pseudo-data were treated as coming from just one ``experiment'', but in practice both ATLAS and CMS would be sources of input data for fitting.

\subsection{Updating PDFs with $A_{FB}$ using the full mass range }

First, we will update PDFs using full mass range $A_{FB}$. 
It is expected that the PDF-induced uncertainty on $A_{FB}(M_{\ell\ell})$ will be reduced after updating the original PDFs with the inclusion of  the pseudo-data. Note that the pseudo-data and theory prediction are generated 
by the same CT14HERA2 PDFs.
If the correlation between the $\effstw$ and the PDF updating is negligible, 
we expect no changes in the central value of $A_{FB}$ as predicted by the PDFs after updating, compared to that given in the original theory prediction.  

\begin{figure}[!hbt]
\begin{center}
\includegraphics[width=0.4\textwidth]{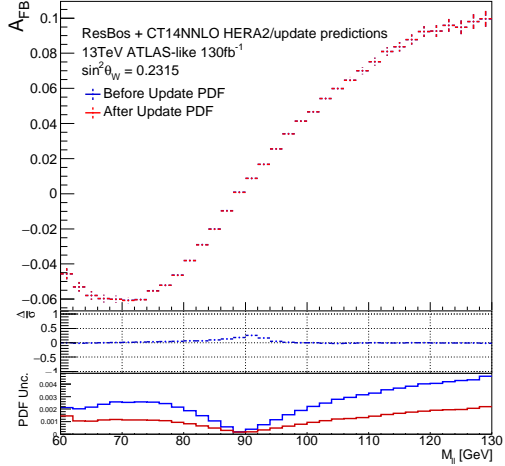}
\caption{\small 
The predicted $A_{FB}$ distributions ($\effstw=0.2315$) and the associated PDF-induced  uncertainties, before 
and after the PDF updating by using the full mass range of the pseudo-data ($\effstw=0.2324$). Only $CC$ events are considered.
The $\Delta/\sigma$ in the middle panel is defined as ($A_{FB}$[before] - $A_{FB}$[after])/$\sigma$, where $\sigma$ is the statistical 
uncertainty of the samples in that bin.
The bottom panel shows the magnitude of the PDF-induced
uncertainty of $A_{FB}$,  predicted by the CT14HERA2 error PDFs, at the 68\% CL., before and after updating the PDFs.
}
\label{fig:stw2324_fullAFBCC}
\end{center}
\end{figure} 

\begin{figure}[!hbt]
\begin{center}
\includegraphics[width=0.4\textwidth]{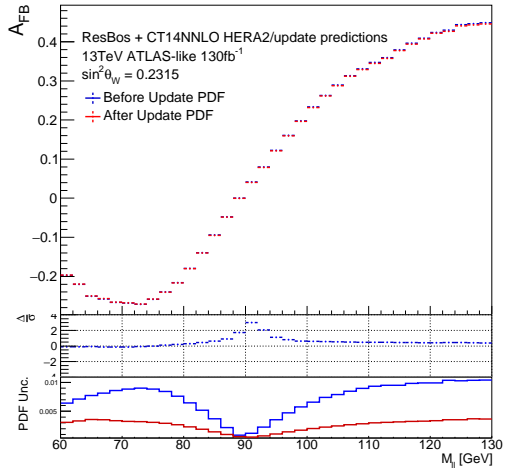}
\caption{\small Similar to Fig.~\ref{fig:stw2324_fullAFBCC}, but for $CF$ events only. 
}
\label{fig:stw2324_fullAFBCF}
\end{center}
\end{figure} 

The predicted $A_{FB}$ distributions ($\effstw=0.2315$) and the associated PDF-induced  uncertainties, before 
and after the PDF updating by using the full mass range of the pseudo-data ($\effstw=0.2324$), are depicted in 
Figs.~\ref{fig:stw2324_fullAFBCC} and ~\ref{fig:stw2324_fullAFBCF} for the $CC$ and $CF$ event samples, respectively.
As shown in the bottom panels of the figures, the PDF-induced uncertainties on the predicted $A_{FB}$ are significantly reduced after 
the updating procedure. This finding is consistent with the conclusion of Ref.~\cite{ArieMassConstrain}.
However, as also shown in the middle panels of the figures, the central values of $A_{FB}$ differ before and after the updating, particularly at the $Z$ pole region. 
The difference is more significant in the $CF$ than $CC$ events. 
When the $\effstw$ of the pseudo-data has a value different from its value in the theory predictions, the 
existing PDF model ({\it i.e.,} CT14HERA2 PDFs, in this study) no longer describes the new data in 
a consistent way. As a result, the PDF updating procedure would forcibly convert this bias, which originated from a different value of 
$\effstw$, into an updated central PDF set. 
The averaged $A_{FB}$ values at the $Z$-pole region in the pseudo-data and theory predictions, before and after PDF updating, are numerically presented in 
Tables~\ref{tab:stw2324_fullAFBCC_updateAFB}, ~\ref{tab:stw2324_fullAFBCF_updateAFB} and ~\ref{tab:stw2324_fullAFBCCCF_updateAFB}, 
 for the $CC$, $CF$ and $CC+CF$ events.
The $CF$ events have higher sensitivity 
to the $A_{FB}$. 
For example, as given in the third line of Table~\ref{tab:stw2324_fullAFBCF_updateAFB}, 
the PDF uncertainty can be decreased from 0.00118 to 0.00055, a reduction of more than 50\%.  Meanwhile, the bias on $A_{FB}$, originating from the PDF updating, can be as large as 
$\Delta=-0.00108$, as shown in the same line of the table. 
As we pointed out, there should be no difference in the central value of $A_{FB}$ before and after an unbiased updating, because 
the pseudo-data and theory templates are generated with same PDF sets.
This bias, which is larger than the statistical uncertainty shown in the third column, indicates that much of the effects of the shift in $\effstw$ have been absorbed into the updated PDFs.

\begin{table*}[hbt]
{\tiny
\begin{center}
\begin{tabular}{l|c|c|c}
\hline \hline
Update using $CC$ with full $A_{FB}$ & average $A_{FB}$ at $Z$ pole & PDF uncertainty & Statistical uncertainty \\
\hline
pseudo-data $\effstw = 0.2324$ & & & \\ 
~~~~ in $CC$ events & 0.00825 & - & 0.00008 \\
~~~~ in $CF$ events & 0.03983 & - & 0.00017 \\
~~~~ in $CC+CF$ events & 0.01368 & - & 0.00007 \\
\hline
theory prediction in $CC$ events, $\effstw = 0.2315$ & & & \\
~~~~ before update & 0.00873 & 0.00038 & 0.00008 \\
~~~~ after update & 0.00867 & 0.00019 & 0.00008 \\
~~~~ $\Delta$[after – before] & -0.00006 & - & - \\
\hline
theory prediction in $CF$ events, $\effstw = 0.2315$ & & & \\
~~~~ before update & 0.04220 & 0.00118 & 0.00017 \\
~~~~ after update & 0.04201 & 0.00092 & 0.00017 \\
~~~~ $\Delta$[after – before] & -0.00019 & - & - \\
\hline
theory prediction in $CC+CF$ events, $\effstw = 0.2315$ & & & \\
~~~~ before update & 0.01449 & 0.00053 & 0.00007 \\
~~~~ after update & 0.01440 & 0.00031 & 0.00007 \\
~~~~ $\Delta$[after – before] & -0.00009 & - & - \\
\hline \hline
\end{tabular}
\caption{\scriptsize Average $A_{FB}$ at $Z$ pole region in the pseudo-data and theory predictions. 
The PDF updating is done using the full mass range $A_{FB}$ spectrum from the $CC$ events of pseudo-data ($\stw=0.2324$).
Statistical uncertainty corresponds to the data sample with an integrated luminosity of 130 fb$^{-1}$. }
\label{tab:stw2324_fullAFBCC_updateAFB}
\end{center}
}
\end{table*} 

\begin{table*}[hbt]
{\tiny
\begin{center}
\begin{tabular}{l|c|c|c}
\hline \hline
Update using $CF$ with full $A_{FB}$ & average $A_{FB}$ at $Z$ pole & PDF uncertainty & Statistical uncertainty \\
\hline
pseudo-data $\effstw = 0.2324$ & & & \\ 
~~~~ in $CC$ events & 0.00825 & - & 0.00008 \\
~~~~ in $CF$ events & 0.03983 & - & 0.00017 \\
~~~~ in $CC+CF$ events & 0.01368 & - & 0.00007 \\
\hline
theory prediction in $CC$ events, $\effstw = 0.2315$ & & & \\
~~~~ before update & 0.00873 & 0.00038 & 0.00008 \\
~~~~ after update & 0.00856 & 0.00026 & 0.00008 \\
~~~~ $\Delta$[after – before] & -0.00017 & - & - \\
\hline
theory prediction in $CF$ events, $\effstw = 0.2315$ & & & \\
~~~~ before update & 0.04220 & 0.00118 & 0.00017 \\
~~~~ after update & 0.04112 & 0.00055 & 0.00017 \\
~~~~ $\Delta$[after – before] & -0.00108 & - & -\\
\hline
theory prediction in $CC+CF$ events, $\effstw = 0.2315$ & & & \\
~~~~ before update & 0.01449 & 0.00053 & 0.00007\\
~~~~ after update & 0.01416 & 0.00032 & 0.00007 \\
~~~~ $\Delta$[after – before] & -0.00033 & - & - \\
\hline \hline
\end{tabular}
\caption{\scriptsize Average $A_{FB}$ at $Z$ pole region in the pseudo-data and theory predictions. 
The PDF updating is done using the full mass range $A_{FB}$ spectrum from the $CF$ events of pseudo-data ($\effstw=0.2324$).
Statistical uncertainty corresponds to the data sample with an integrated luminosity of 130 fb$^{-1}$. }
\label{tab:stw2324_fullAFBCF_updateAFB}
\end{center}
}
\end{table*} 

\begin{table*}[hbt]
{\tiny
\begin{center}
\begin{tabular}{l|c|c|c}
\hline \hline
Update using $CC$ + $CF$ with full $A_{FB}$ & average $A_{FB}$ at $Z$ pole & PDF uncertainty & Statistical uncertainty \\
\hline
pseudo-data $\effstw = 0.2324$ & & & \\ 
~~~~ in $CC$ events & 0.00825 & - & 0.00008 \\
~~~~ in $CF$ events & 0.03983 & - & 0.00017 \\
~~~~ in $CC+CF$ events & 0.01368 & - & 0.00007 \\
\hline
theory prediction in $CC$ events, $\effstw = 0.2315$ & & & \\
~~~~ before update & 0.00873 & 0.00038 & 0.00008 \\
~~~~ after update & 0.00853 & 0.00018 & 0.00008 \\
~~~~ $\Delta$[after – before] & -0.00020 & - & - \\
\hline
theory prediction in $CF$ events, $\effstw = 0.2315$ & & & \\
~~~~ before update & 0.04220 & 0.00118 & 0.00017 \\
~~~~ after update & 0.04105 & 0.00054 & 0.00017 \\
~~~~ $\Delta$[after – before] & -0.00115 & -& -\\
\hline
theory prediction in $CC+CF$ events, $\effstw = 0.2315$ & & & \\
~~~~ before update & 0.01449 & 0.00053 & 0.00007\\
~~~~ after update & 0.01411 & 0.00025 & 0.00007 \\
~~~~ $\Delta$[after – before] & -0.00038 & & \\
\hline \hline
\end{tabular}
\caption{\scriptsize Average $A_{FB}$ at $Z$ pole region in the pseudo-data (with $\stw=0.2324$) and theory predictions (with $\stw=0.2315$). 
The PDF updating is done using the full mass range $A_{FB},$ from the  $CC$, $CF$ or $CC+CF$ events of pseudo-data. 
Statistical uncertainty corresponds to the data sample with an integrated luminosity of 130 fb$^{-1}$. }
\label{tab:stw2324_fullAFBCCCF_updateAFB}
\end{center}
}
\end{table*}

To estimate the impact on the determination of $\effstw$ in the $Z$-pole mass region, we express the average $A_{FB}$ approximately 
as a linear function of $\effstw$ in this region, written as 
\begin{eqnarray}
A_{FB} \simeq k \cdot \effstw + b \, ,
\end{eqnarray}
where the values of the parameters $k$ and $b$ are listed in Table~\ref{tab:afb_stw_function}, for $CC$, $CF$ and $CC+CF$ event samples, respectively.  

\begin{table}[hbt]
\begin{center}
\begin{tabular}{|l|c|c|}
\hline \hline
& slope factor $k$ & offset factor $b$ \\
\hline
$CC$ events: & -0.531 & 0.132 \\
\hline
$CF$ events: & -2.512 & 0.623 \\
\hline
$CC+CF$ events: & -1.110 & 0.275 \\
\hline \hline
\end{tabular}
\caption{\small Simple linear functions between $\effstw$ and the observed $A_{FB}$ around 
$Z$ pole, predicted by ResBos with CT14NNLO PDFs, for the $CC$, $CF$ and $CC+CF$ event samples, respectively.}
\label{tab:afb_stw_function}
\end{center}
\end{table}

One could roughly estimate the bias and the PDF-induced uncertainty on the determination of $\effstw$, derived from the biased $A_{FB}$ measurement, using the 
following simplified relation:
\begin{eqnarray}
\Delta \effstw = \Delta A_{FB}/k \,.
\end{eqnarray}
From the above equation and  Table~\ref{tab:stw2324_fullAFBCCCF_updateAFB}, we obtain the results listed in Table~\ref{tab:fullmassAFB_stw}.
\begin{table}[H]
\begin{center}
\begin{tabular}{|l|c|c|}
\hline \hline
Update PDF by using & Potential bias & PDF uncertainty\\ 
full mass range $A_{FB}$ & on $\effstw$ & on $\effstw$ \\
\hline
$CC$ events: & 0.00038 & 0.00033 \\
\hline
$CF$ events: & 0.00046 & 0.00021 \\
\hline \hline
\end{tabular}
\caption{\small The bias and the PDF-induced uncertainty on $\effstw$, after updating the PDFs with the full mass range of the  $A_{FB}$ pseudo-data ($\effstw=0.2324$), for the $CC$ and $CF$ event samples, respectively.
The PDF uncertainties are given at the 68\% C.L.
}
\label{tab:fullmassAFB_stw}
\end{center}
\end{table} 

It can be seen that the bias on $\effstw$ determined from the biased $A_{FB}$ after the PDF updating is much larger than the PDF-induced uncertainty itself, 
especially in the $CF$ event sample which is more sensitive to $\effstw$ than the $CC$ event sample.
Of course, this bias depends on the difference between $\effstw$ values in the pseudo-data and the original theory prediction. And in this work, it is intentionally 
set to an exaggeratedly large difference of 0.0009 for illustration, which is 3 times the uncertainty obtained from the best hadron collider measurements.
A smaller difference between the $\effstw$ value of the pseudo-data and the world average value would surely lead to a smaller bias in the $A_{FB}$ measurement 
after the PDF updating procedure. Nevertheless, this part of our study clearly demonstrates the fact that using the full mass spectrum of the $A_{FB}$ data to update 
the existing PDFs will introduce bias in the determination of $\effstw$ at the $Z$-pole mass region.
With more data collected at the future high luminosity LHC, the weak mixing angle can be determined more precisely, and  the  $\effstw$ measurements with different lepton 
final states of DY processes at the ALTAS, CMS and LHCb experiments should be considered as separate measurements.
Occasionally, one might expect some individual $\effstw$ measurements to exhibit significant deviations from the nominal world average 
value. In such circumstances, the potential bias on the $\effstw$ extraction, induced by updating PDFs with the $A_{FB}$ measurement spanning the full mass range, 
from 60 GeV to 130 GeV, should be seriously considered.

\begin{figure}[hbt]
\begin{center}
\includegraphics[width=0.4\textwidth]{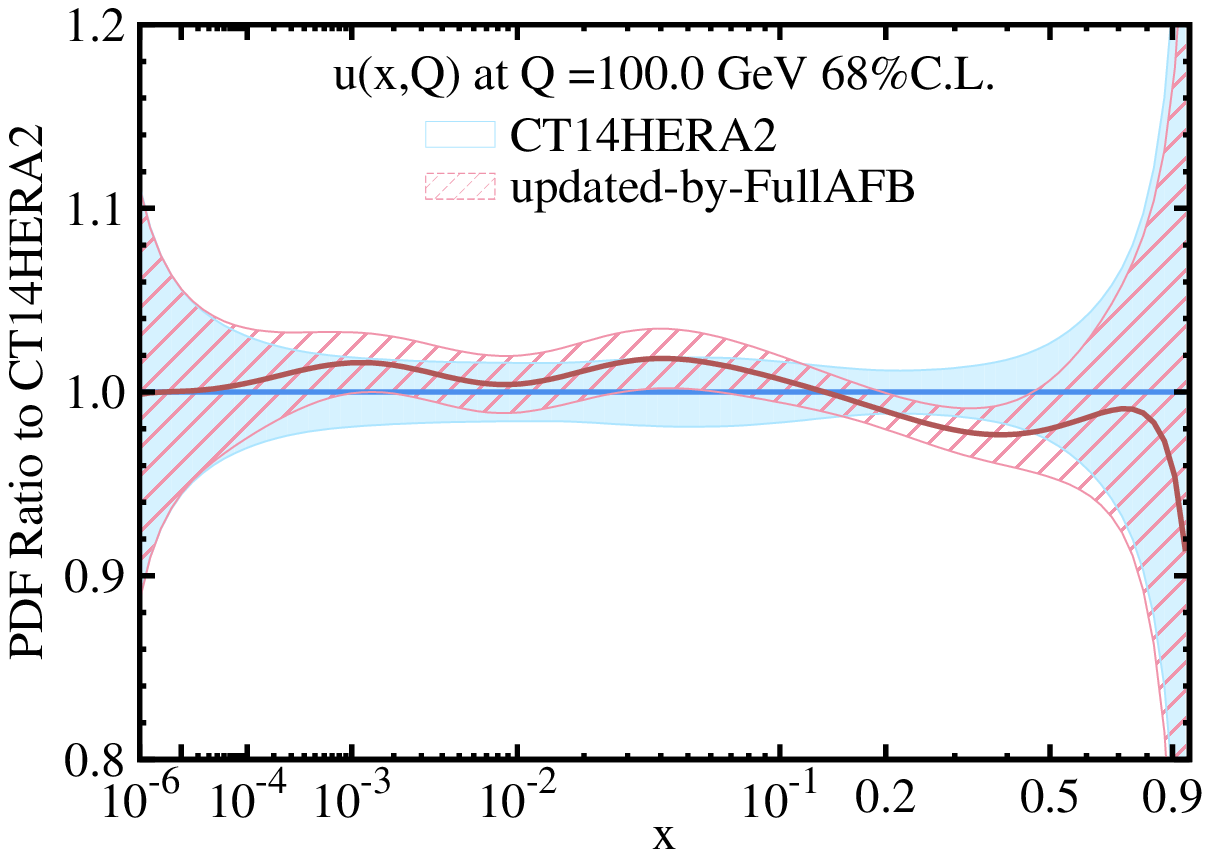}
\includegraphics[width=0.4\textwidth]{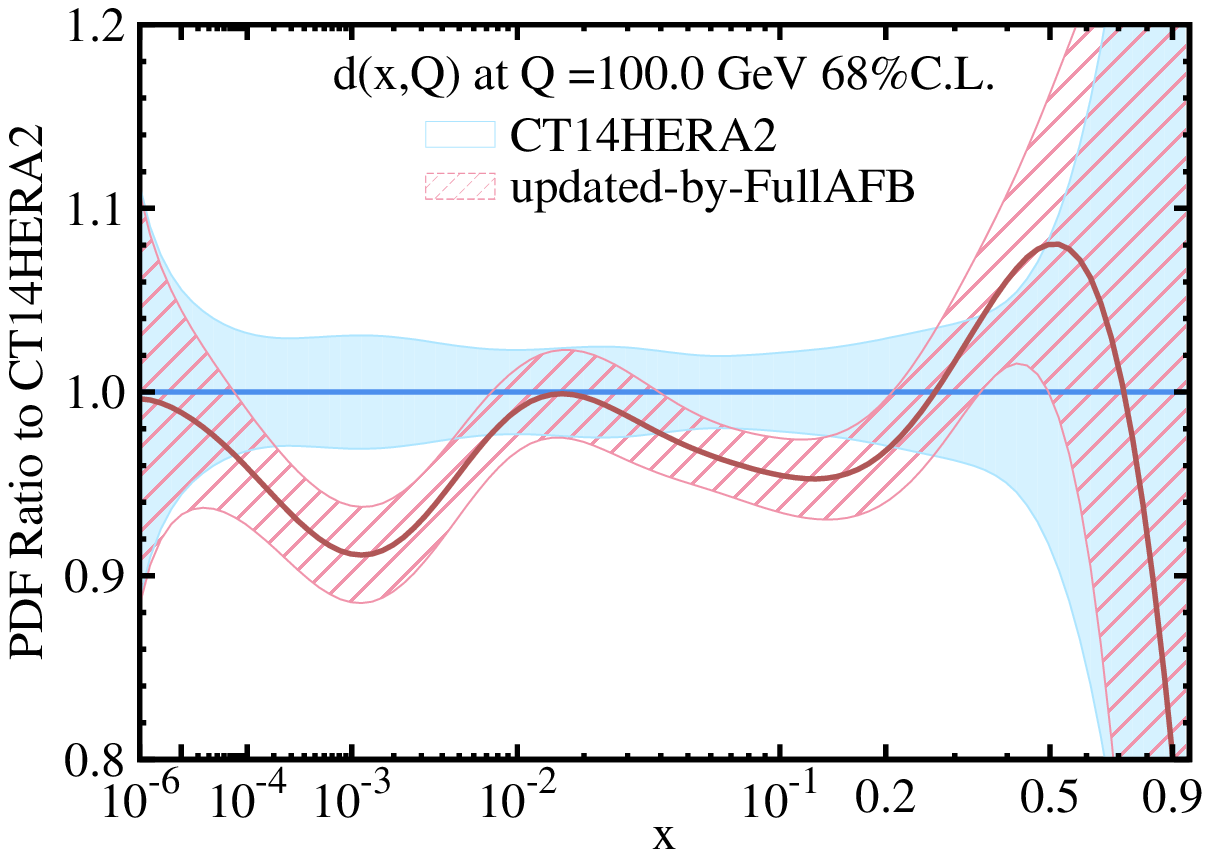}
\caption{\small Ratios of the central value and uncertainty to the CT14HERA2 central value of the $u$ and $d$ PDFs, before and after the PDF updating. The blue band corresponds to the uncertainty before updating and the 
red band is after updating. Full mass range of the $A_{FB}$ pseudo-data sample  (with $\stw=0.2324$) is used for the updating.
}
\label{fig:fullmassPDF}
\end{center}
\end{figure}

The bias incurred by updating the PDFs using the full mass spectrum can also be seen by looking directly at the PDFs of the quarks and gluons themselves.
Fig.~\ref{fig:fullmassPDF} depicts the comparison of $u$ and $d$ quark PDFs before and 
after the updating. 
With an unbiased updating procedure, the central PDF values of the two PDF sets (before and after the PDF updating)  should be unchanged, while the updated PDF uncertainties are expected to be 
reduced after the inclusion of the new pseudo-data. This feature, however, is not confirmed in  Fig.~\ref{fig:fullmassPDF}.  Again, this displays how the 
biased updated PDFs have been changed in order to compensate for the effects of the shifted $\effstw$ in the pseudo-data.

\subsection{Updating PDFs using sideband $A_{FB}$ data only}

As shown in FIG.~\ref{fig:AFBfullmass_comparestw}, the $A_{FB}$ asymmetry is more sensitive to $\effstw$ when $M_{\ell\ell}$ is around the $Z$-pole mass, 
while the PDF-induced uncertainty becomes more significant when $M_{\ell\ell}$ is outside the $Z$ pole mass window. 
This is because that in the $Z$ pole region, the asymmetry is proportional to both the vector and axial-vector couplings of the $Z$ boson to the fermions and 
is numerically close to 0. And since only the vector coupling of the $Z$ boson depends on the weak mixing angle, the information on $\effstw$ predominantly 
comes from the $A_{FB}$ in the vicinity of the Z-boson pole. While away from the Z-boson mass pole, the asymmetry results from the interference of 
the axial vector Z coupling and vector photon coupling and depends upon the PDFs. 
On the other hand, the sensitivity of constraining the PDFs via a measurement of $A_{FB}$ depends on the value of the asymmetry (see Appendix A). 
Consequently, the $A_{FB}$-to-PDF sensitivity is suppressed in the $Z$ pole region where the value of the asymmetry is close to zero, and is enhanced outside the $Z$ pole mass window with magnified $A_{FB}$ value.
This observation suggests that we could separate the $A_{FB}$ distribution into $Z$ pole region and sideband region, and use them for $\effstw$ determination and PDF updating procedure, respectively. This procedure could reduce the correlation, and keep most of the sensitivities.

To confirm this, we generate and use 
a new pseudo-data sample with an even more different value of $\effstw$ (as = 0.2345) in this section, {\it i.e.,} the difference between 
$\effstw$ values in the pseudo-data and the original theory templates is 10 times the best precision at hadron colliders. 
When this new pseudo-data sample was generated, $Z$ pole events with $M_{\ell\ell}$ from 80 to 100 GeV were explicitly excluded. 
Following the same analysis procedures as discussed in the previous section, 
we obtain the numerical results listed in Tables~\ref{tab:stw2345_sidebandAFBCC_updateAFB}, ~\ref{tab:stw2345_sidebandAFBCF_updateAFB} and 
~\ref{tab:stw2345_sidebandAFBCCCF_updateAFB}, which summarize the impact of $CC$, $CF$ and $CC+CF$ events, respectively. 

\begin{table*}[hbt]
{\tiny
\begin{center}
\begin{tabular}{l|c|c|c}
\hline \hline
  Update using $CC$ with sideband $A_{FB}$ & average $A_{FB}$ at $Z$ pole & PDF uncertainty & Statistical uncertainty \\
\hline
 pseudo-data $\effstw = 0.2345$ & & & \\ 
 ~~~~ in $CC$ events &  0.00714 & - & 0.00008 \\
 ~~~~ in $CF$ events &  0.03490 & - & 0.00017 \\
 ~~~~ in $CC+CF$ events & 0.01192 & - & 0.00007 \\
\hline
 theory prediction in $CC$ events, $\effstw = 0.2315$ & & & \\
 ~~~~ before update & 0.00873 & 0.00038 & 0.00008 \\
 ~~~~ after update & 0.00872 & 0.00024 & 0.00008 \\
 ~~~~ $\Delta$[after – before] &  -0.00001 & - & - \\
\hline
  theory prediction in $CF$ events, $\effstw = 0.2315$ & & & \\
 ~~~~ before update & 0.04220 & 0.00118 & 0.00017 \\
 ~~~~ after update & 0.04218 & 0.00098 & 0.00017 \\
 ~~~~ $\Delta$[after – before] & -0.00002 & - & - \\
\hline
  theory prediction in $CC+CF$ events, $\effstw = 0.2315$ & & & \\
 ~~~~ before update & 0.01449 & 0.00053 & 0.00007 \\
 ~~~~ after update & 0.01448 & 0.00036 & 0.00007 \\
 ~~~~ $\Delta$[after – before] & -0.00001 & - & - \\
\hline \hline
\end{tabular}
\caption{\scriptsize Average $A_{FB}$ at $Z$ pole region in the pseudo-data and theory predictions. 
The PDF updating is done using the sideband $A_{FB}$ spectrum from the $CC$ events of pseudo-data ($\effstw=0.2345$).
Statistical uncertainty corresponds to the data sample with an integrated luminosity of 130 fb$^{-1}$. }
\label{tab:stw2345_sidebandAFBCC_updateAFB}
\end{center}
}
\end{table*} 

\begin{table*}[hbt]
{\tiny
\begin{center}
\begin{tabular}{l|c|c|c}
\hline \hline
  Update using $CF$ with sideband $A_{FB}$ & average $A_{FB}$ at $Z$ pole & PDF uncertainty & Statistical uncertainty \\
\hline
 pseudo-data $\effstw = 0.2345$ & & & \\ 
 ~~~~ in $CC$ events & 0.00714 & - & 0.00008 \\
 ~~~~ in $CF$ events & 0.03490 & - & 0.00017 \\
 ~~~~ in $CC+CF$ events & 0.01192 & - & 0.00007 \\
\hline
 theory prediction in $CC$ events, $\effstw = 0.2315$ & & & \\
~~~~ before update & 0.00873 & 0.00038 & 0.00008 \\
~~~~ after update & 0.00868 & 0.00027 & 0.00008 \\
~~~~ $\Delta$[after – before] &  -0.00005 & - & - \\
\hline
theory prediction in $CF$ events, $\effstw = 0.2315$ & & & \\
~~~~ before update & 0.04220 & 0.00118 & 0.00017 \\
~~~~ after update & 0.04172 & 0.00073 & 0.00017 \\
~~~~ $\Delta$[after – before] & -0.00048 & - & - \\
\hline
theory prediction in $CC+CF$ events, $\effstw = 0.2315$ & & & \\
~~~~ before update & 0.01449 & 0.00053 & 0.00007 \\
~~~~ after update & 0.01437 & 0.00036 & 0.00007 \\
~~~~ $\Delta$[after – before] & -0.00012 & - & - \\
\hline \hline
\end{tabular}
\caption{\scriptsize Average $A_{FB}$ at $Z$ pole region in the pseudo-data and theory predictions. 
The PDF updating is done using the sideband $A_{FB}$ spectrum from the $CF$ events of pseudo-data ($\effstw=0.2345$).
Statistical uncertainty corresponds to the data sample with an integrated luminosity of 130 fb$^{-1}$. }
\label{tab:stw2345_sidebandAFBCF_updateAFB}
\end{center}
}
\end{table*} 
 \begin{table*}[hbt]
{\tiny
\begin{center}
\begin{tabular}{l|c|c|c}
\hline \hline
  Update using $CC$ + $CF$ with sideband $A_{FB}$ & average $A_{FB}$ at $Z$ pole & PDF uncertainty & Statistical uncertainty \\
\hline
 pseudo-data $\effstw = 0.2345$ & & & \\ 
 ~~~~ in $CC$ events & 0.00714 & - & 0.00008 \\
 ~~~~ in $CF$ events &  0.03490 & - & 0.00017 \\
 ~~~~ in $CC+CF$ events & 0.01192 & - & 0.00007 \\
\hline
 theory prediction in $CC$ events, $\effstw= 0.2315$ & & & \\
~~~~ before update & 0.00873 & 0.00038 & 0.00008 \\
~~~~ after update & 0.00868 & 0.00022 &  0.00008 \\
~~~~ $\Delta$[after – before] & -0.00005 & - & - \\
\hline
theory prediction in $CF$ events, $\effstw = 0.2315$ & & & \\
~~~~ before update & 0.04220 & 0.00118 & 0.00017 \\
~~~~ after update & 0.04173 & 0.00072 & 0.00017 \\
~~~~ $\Delta$[after – before] & -0.00047 & -& -\\
\hline
theory prediction in $CC+CF$ events, $\effstw = 0.2315$ & & & \\
~~~~ before update & 0.01449 & 0.00053 & 0.00007\\
~~~~ after update & 0.01437 & 0.00032 & 0.00007 \\
~~~~ $\Delta$[after – before] & -0.00012 & & \\
\hline \hline
\end{tabular}
\caption{\scriptsize Average $A_{FB}$ at $Z$ pole region in the pseudo-data and theory predictions. 
The PDF updating is done using the sideband $A_{FB}$ spectra from both the $CC$ and $CF$ events of pseudo-data ($\effstw=0.2345$).
Statistical uncertainty corresponds to the data sample with an integrated luminosity of 130 fb$^{-1}$. }
\label{tab:stw2345_sidebandAFBCCCF_updateAFB}
\end{center}
}
\end{table*}  

\begin{figure}[hbt]
\begin{center}
\includegraphics[width=0.4\textwidth]{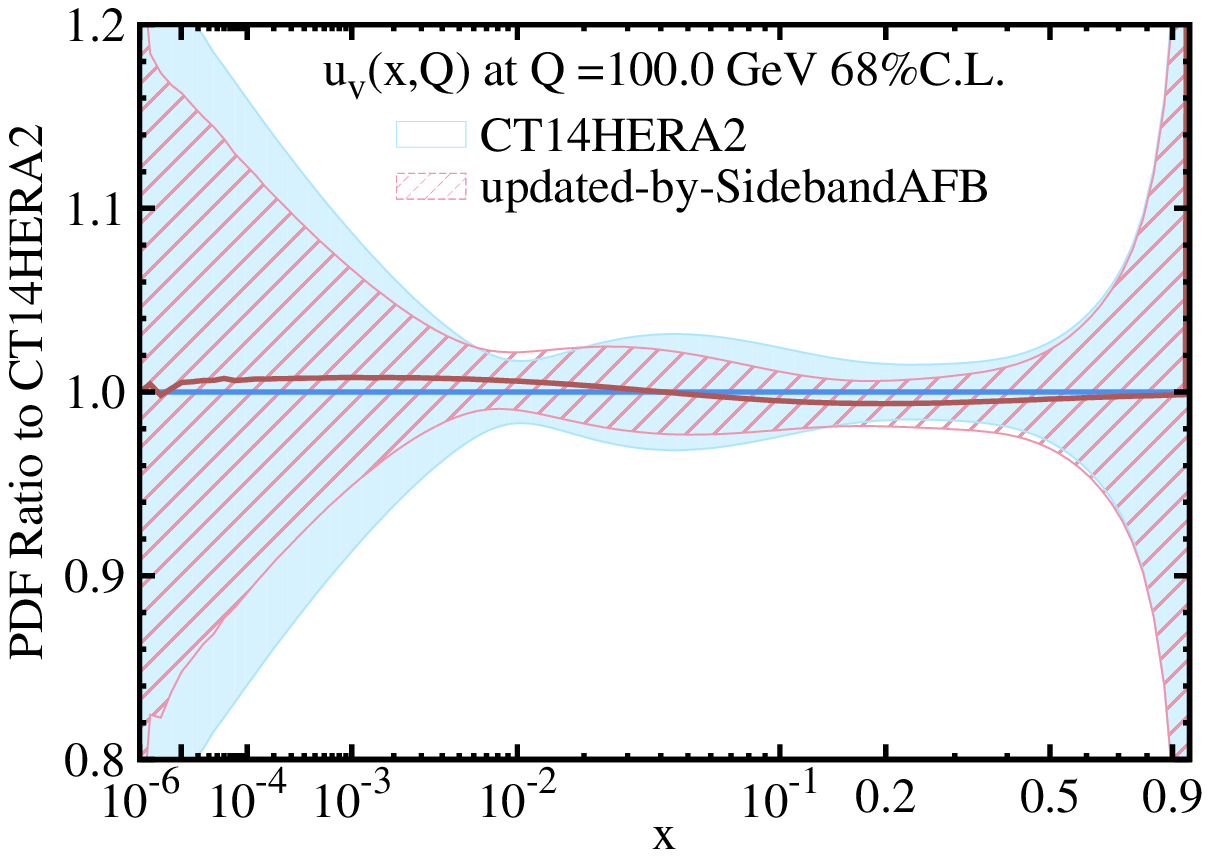}
\includegraphics[width=0.4\textwidth]{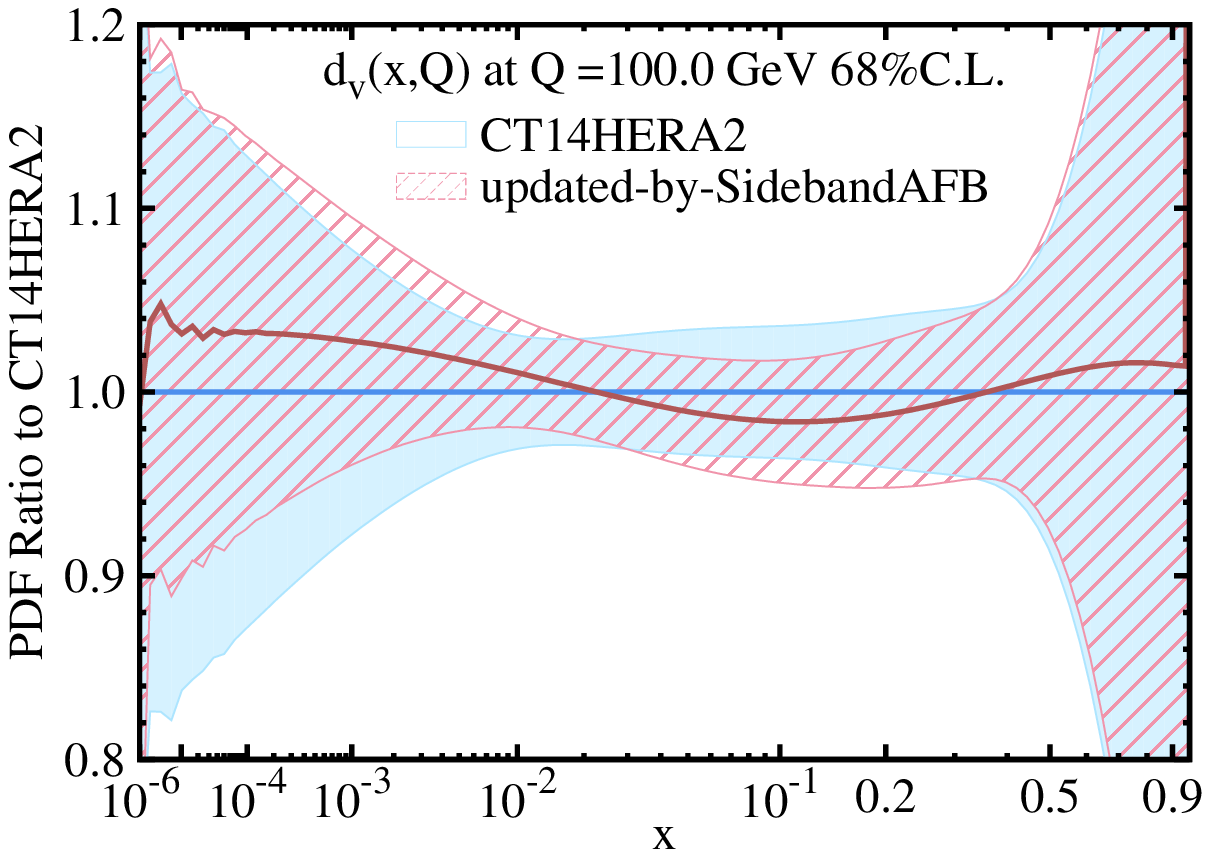}
\includegraphics[width=0.4\textwidth]{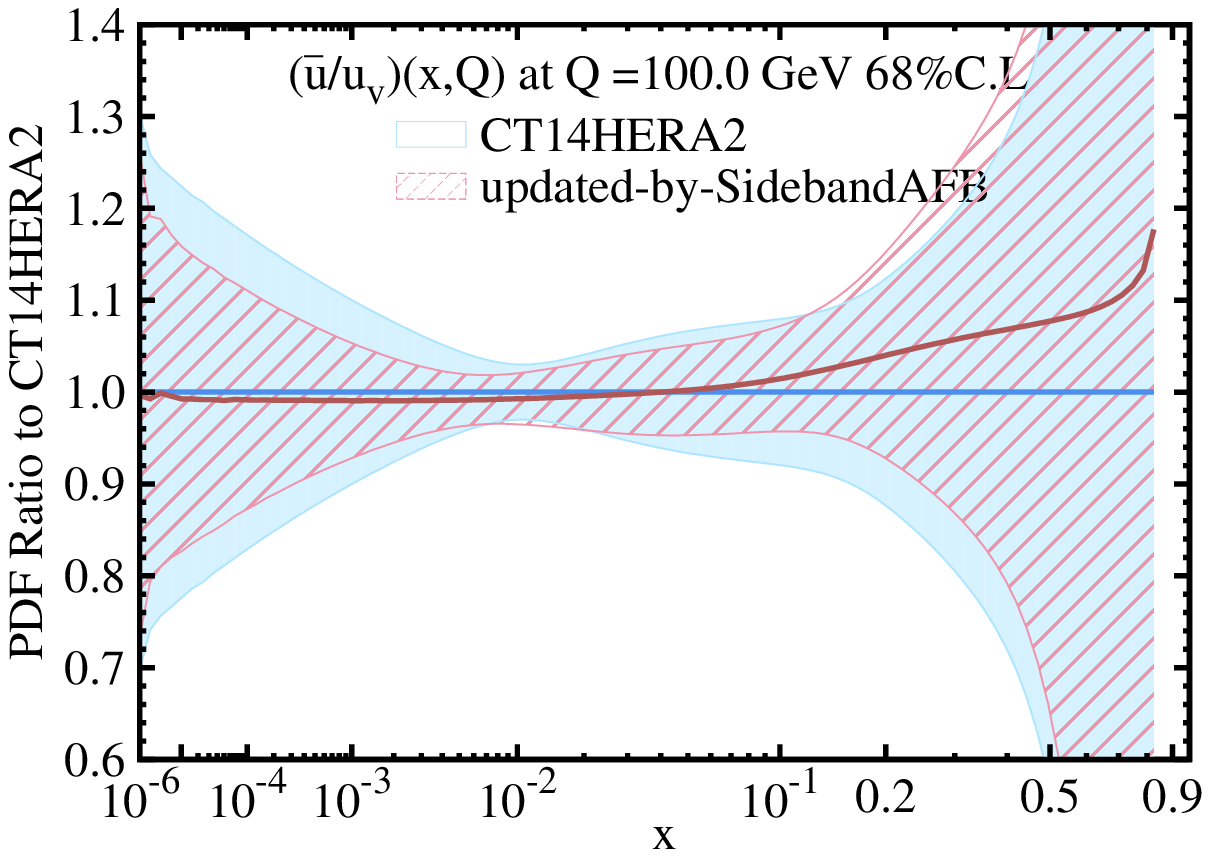}
\includegraphics[width=0.4\textwidth]{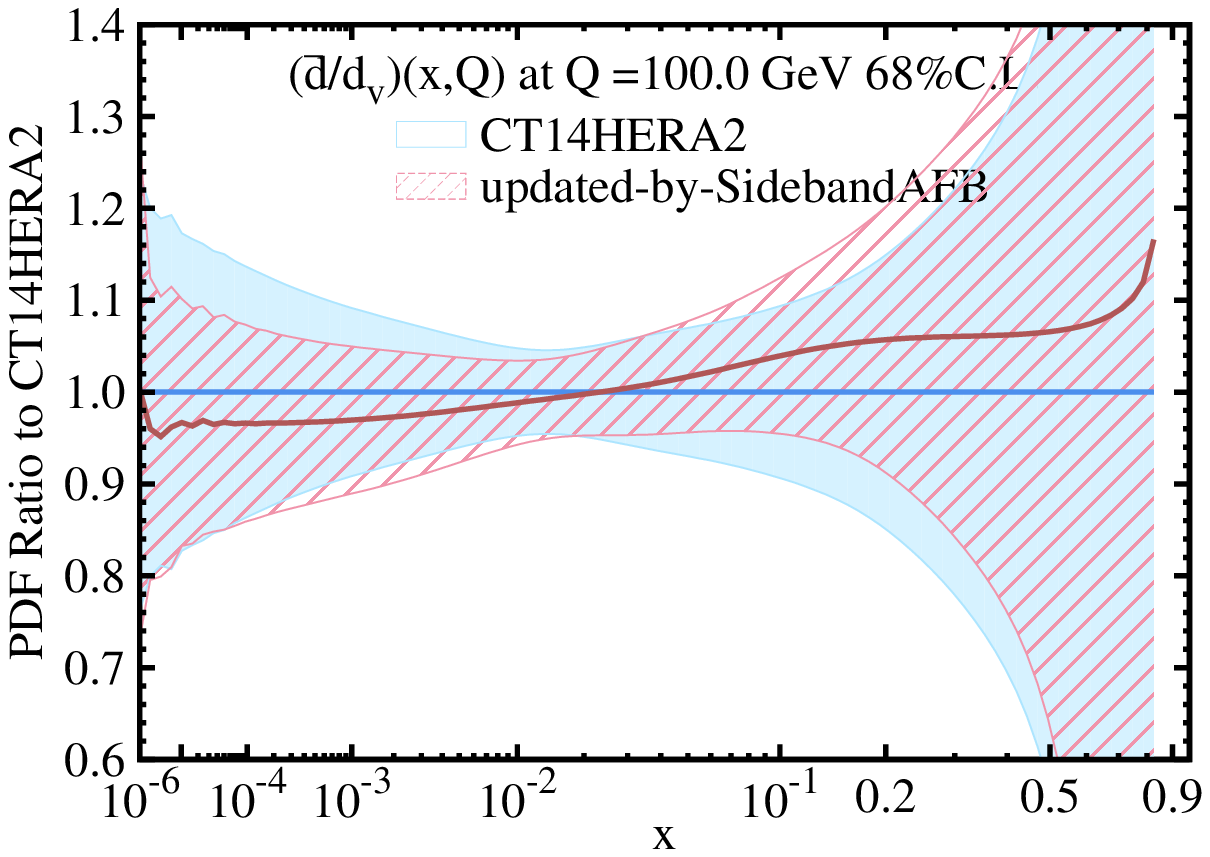}
\caption{\small Ratios of the central value and uncertainty to the CT14HERA2 central value of the $u_v$, $d_v$, ${\bar u}/{u_v}$ and ${\bar d}/{d_v}$ PDFs, before and after the PDF updating. The blue band corresponds to the uncertainty before updating and the 
red band is after updating. Sideband mass range of the $A_{FB}$ pseudo-data sample  ($\effstw=0.2345$) is used for the updating.
}
\label{fig:sidebandmassPDF}
\end{center}
\end{figure}

Since the inclusive production rate of the $Z$ boson is dominated by the contribution from the $Z$-pole mass window, the constraint on the PDF 
uncertainty obtained from using only the sideband $A_{FB}$ data sample is not as statistically powerful as that using the full mass range $A_{FB}$ data sample. 
For example, comparing the sideband result (in  Table~\ref{tab:stw2345_sidebandAFBCCCF_updateAFB}) to the full mass range 
result (in  Table~\ref{tab:stw2324_fullAFBCCCF_updateAFB}), we find that the PDF uncertainty only reduces to 0.00072 for sideband updating, 
compared to 0.00054 for full mass range updating, in the case of using the most sensitive $CF$ event sample. But, on the other hand, 
the bias on the average $A_{FB}$ in the $Z$-pole mass window is much smaller in the sideband updating (with $\Delta=-0.00047$ and  $\effstw=0.2345$) 
than that in the full mass range updating  (with $\Delta=-0.00115$ and $\effstw=0.2324$), as listed in the same tables.
Furthermore, in contrast to the strong variation observed in Fig.~\ref{fig:fullmassPDF}, we find 
much less bias on various parton flavor PDFs when using only the 
sideband $A_{FB}$ data to update the PDFs, as shown in Fig.~\ref{fig:sidebandmassPDF}.

By using the numbers from  Table~\ref{tab:stw2345_sidebandAFBCCCF_updateAFB}, the impact of updating PDFs with the sideband $A_{FB}$ data on 
the determination of $\effstw$ can be summarized in  Table~\ref{tab:sidebandAFB_stw}. 
In comparison to the result of using the full mass range $A_{FB}$ data in the updating, cf. Table~\ref{tab:fullmassAFB_stw}, the PDF-induced uncertainty 
from using only the sideband $A_{FB}$ data sample increases by about 20-30\%, but the biases on $\effstw$ diminish dramatically, despite using the much 
larger value of $\effstw=0.2345$ in the present case.
Since the bias introduced by using the full mass range $A_{FB}$ updating is apparently larger than the PDF-induced uncertainties, 
to reduce the bias on $\effstw$ by using the sideband $A_{FB}$ updating should have higher priority than to keep the statistical uncertainty $20\%\sim 30\%$ smaller. 
One should optimize the mass window for a specific measurement to have better balance between bias and sensitivities.

\begin{table}[H]
\begin{center}
\begin{tabular}{|l|c|c|}
\hline \hline
Update PDF using & Potential bias & PDF uncertainty\\ 
sideband $A_{FB}$ & on $\effstw$ & on $\effstw$  \\
\hline
$CC$ events: & 0.00009 & 0.00041 \\
\hline
$CF$ events: & 0.00019 & 0.00029 \\
\hline \hline
\end{tabular}
\caption{\small 
 The bias and the PDF-induced uncertainty on $\stw$, after updating the PDFs with the sideband range of the  $A_{FB}$ pseudo-data (with $\stw=0.2345$), for the $CC$ and $CF$ event samples, respectively.
The PDF uncertainties are given at the 68\% C.L.
}
\label{tab:sidebandAFB_stw}
\end{center}
\end{table}

Considering the fact that the new data might not correspond to a value of $\effstw$ as large as 0.2345, we could conclude that by the end of the LHC Run 2, 
the potential bias of using the sideband $A_{FB}$ data in the PDF updating should be small. However, this does not mean one can ignore it. As we have seen, 
by using the sideband $A_{FB}$ in the PDF updating one can reduce the effects of any potential bias, while not significantly enlarging the total uncertainty 
of the $\effstw$ determination. 
Furthermore, we still strongly suggest keeping the PDF updating as a preliminary method to improve the $\effstw$ measurement.
A final determination of $\effstw$ and its uncertainty estimation can only be reliably provided by a full global analysis, which includes new data sets and allows 
a thorough study on adding new degrees of freedom in the nonperturbative PDF parameters, etc. Experimental results should also be provided in a proper format,  
allowing theorists to replace the preliminary PDF updating method employed in the experimental measurement by a consistent global analysis.

\section{Updating PDFs using lepton charge asymmetry $A_\pm(\eta_\ell)$ in $W$ production}

In this section, we investigate the advantage of using the asymmetry in the rapidity distribution of the charged leptons from $W\rightarrow l\nu$ boson decays, 
produced at the LHC, to update the PDFs. In $pp$ collisions, $W^+$ and $W^-$ have different cross sections, and accordingly an asymmetry can be defined 
as a function of the final state charged lepton rapidity $\eta_\ell$:
\begin{eqnarray}
A_\pm(\eta_\ell) = \frac{N_{W^+}(\eta_\ell) - N_{W^-}(\eta_\ell)}{N_{W^+}(\eta_\ell) + N_{W^-}(\eta_\ell)}.
\end{eqnarray}
This asymmetry is caused by the difference between up and down type quark and their anti-quark distributions in the proton, and thus provides complementary information to $A_{FB}$ 
in constraining the PDFs. 
Although using the $ A_\pm(\eta_\ell)$ as input is essential to many other PDF constraints, it has less impact on the $\effstw$ measurements, compared to using the $A_{FB}$ in the PDF updating. 
In general, the $A_\pm(\eta_\ell)$ is an initial state asymmetry, directly reflcting the difference between $W^+$ and $W^-$ production rates at the LHC, 
and has little dependence on the weak interaction decays.

\begin{figure}[hbt]
\begin{center}
\includegraphics[width=0.4\textwidth]{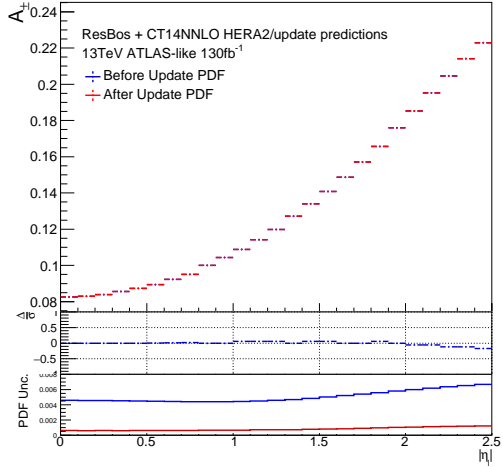}
\caption{\small Theory prediction of $A_\pm(\eta_\ell)$ at the 13 TeV LHC with an integrated luminosity of 130 fb$^{-1}$, as a function of the charged lepton rapidity $\eta_\ell$. 
The $\Delta/\sigma$ in the middle panel is defined as ($A_\pm(\eta_\ell)$[before] - $A_\pm(\eta_\ell)$[after])/$\sigma$, where $\sigma$ is the statistical 
uncertainty of the samples in that bin.
The bottom panel shows the magnitude of the PDF-induced
uncertainty of $A_\pm(\eta_\ell)$,  predicted by the CT14HERA2 error PDFs, at the 68\% CL., before and after updating the PDFs.
}
\label{fig:WasymmetryShape}
\end{center}
\end{figure}

To study the impact of the $A_\pm(\eta_\ell)$ on reducing the PDF-induced uncertainty in the $A_{FB}$ measurement, we generate a set of $W$ boson samples, in which the 
$\effstw$ value is taken to be different from the original theory templates, as done in the previous DY case. 
To model the ATLAS acceptance, the charged leptons (electrons and muons) from the $W$ boson decay are required to have their $|\eta_\ell|< 2.5$. 
Forward electrons are usually removed from the single $W$ production measurement, due to difficulties in controlling the backgrounds 
in the high rapidity region. Both charged leptons and neutrinos are required to have $p_T$ $>$ 25 GeV. A bin size of  0.1 on $|\eta_\ell|$ is used 
in the $A_\pm(\eta_\ell)$ distributions. The $A_\pm(\eta_\ell)$ distributions, together with the PDF-induced uncertainties, before and after the PDF updating, 
are shown in Fig.~\ref{fig:WasymmetryShape}.

The values of the average $A_{FB}$ and their PDF-induced uncertainty after updating PDFs with the simulated lepton charge asymmetry pseudo-data 
are listed in Table~\ref{tab:WasymmetryupdateResult}.
The PDF-induced uncertainty on the average $A_{FB}$ is reduced by $17\%$ for $CC$, and $13\%$ for $CF$ events, 
after updating PDFs with the  $A_\pm(\eta_\ell)$ data. 
The central prediction for $A_{FB}$ does not change after updating PDFs with the  $A_\pm(\eta_\ell)$ data, since there is no 
direct correlation between the value of $\effstw$ and $A_\pm(\eta_\ell)$.
  
\begin{table*}[hbt]
{\tiny
\begin{center}
\begin{tabular}{l|c|c|c|}
\hline \hline
  Update using $A_\pm(\eta_\ell)$ in pseudo-data & average $A_{FB}$ at $Z$ pole & PDF uncertainty & Statistical uncertainty \\
\hline
 theory prediction in $CC$ events, $\effstw = 0.2315$ & & & \\
~~~~ before update & 0.00873 & 0.00038 & 0.00008 \\
~~~~ after update & 0.00873 & 0.00031 & 0.00008 \\
~~~~ $\Delta$[before - after] & $<$0.00001 & & \\
\hline
theory prediction in $CF$ events, $\effstw = 0.2315$ & & & \\
~~~~ before update & 0.04220 & 0.00118 & 0.00017 \\
~~~~ after update & 0.04219 & 0.00103 & 0.00017 \\
~~~~ $\Delta$[before - after]  & 0.00001 & & \\
\hline
theory prediction in $CC+CF$ events, $\effstw = 0.2315$ & & & \\
~~~~ before update & 0.01449 & 0.00053 & 0.00007\\
~~~~ after update & 0.01449 & 0.00044 & 0.00007 \\
~~~~ $\Delta$[before - after]  & $<$0.00001 & & \\
\hline \hline
\end{tabular}
\caption{\scriptsize Average $A_{FB}$ at $Z$ pole region before and after the PDF updating. 
The PDF updating is done using $A_\pm(\eta_\ell)$ from $W$ boson production. 
Statistical uncertainty corresponds to the data sample with an integrated luminosity of 130 fb$^{-1}$. }
\label{tab:WasymmetryupdateResult}
\end{center}
}
\end{table*}  

\begin{figure}[hbt]
\begin{center}
\includegraphics[width=0.4\textwidth]{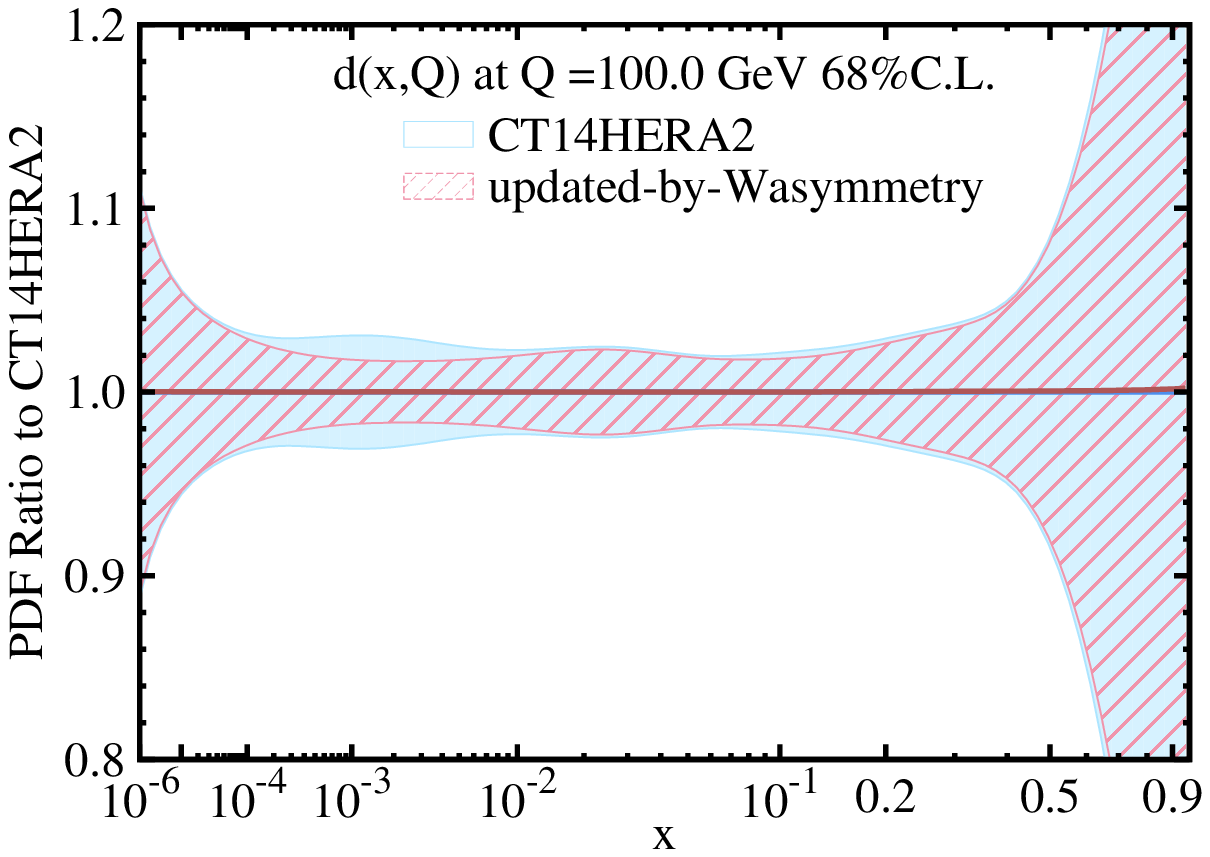}
\includegraphics[width=0.4\textwidth]{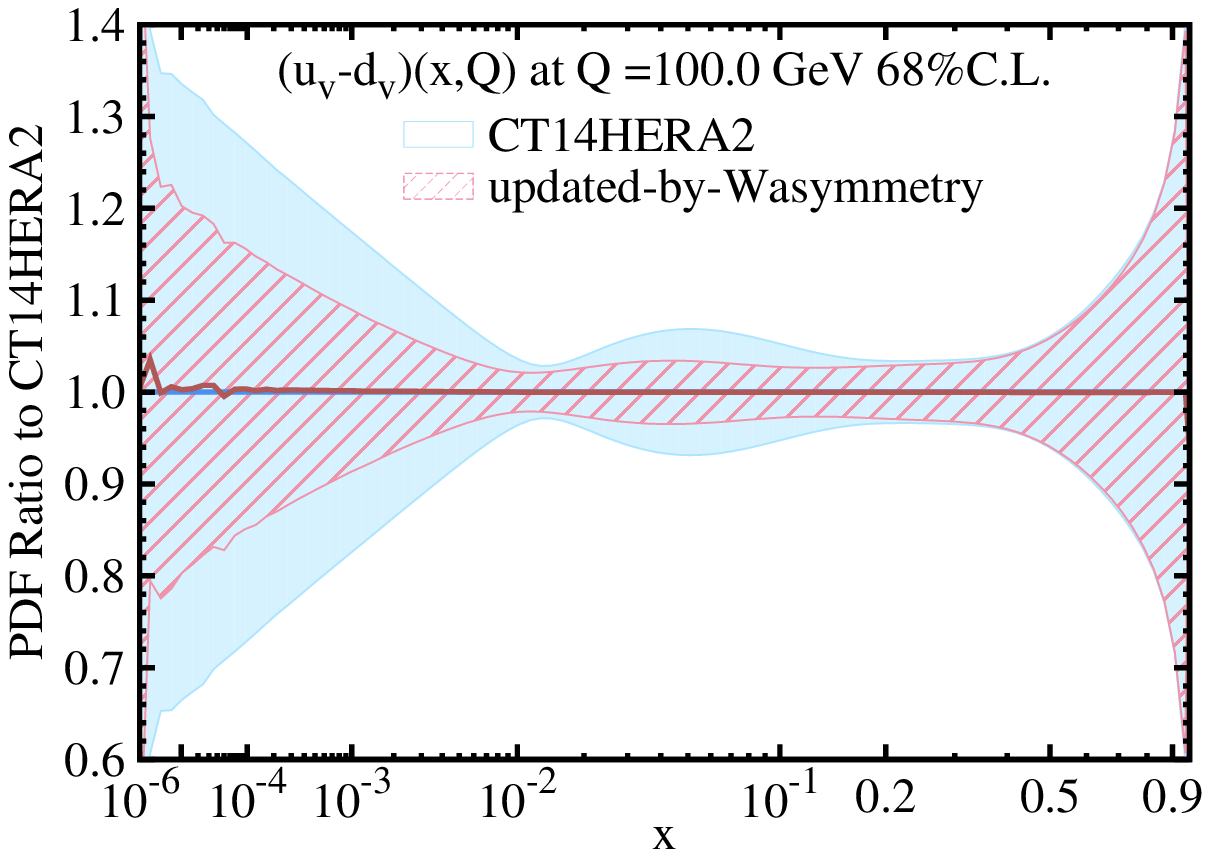}
\includegraphics[width=0.4\textwidth]{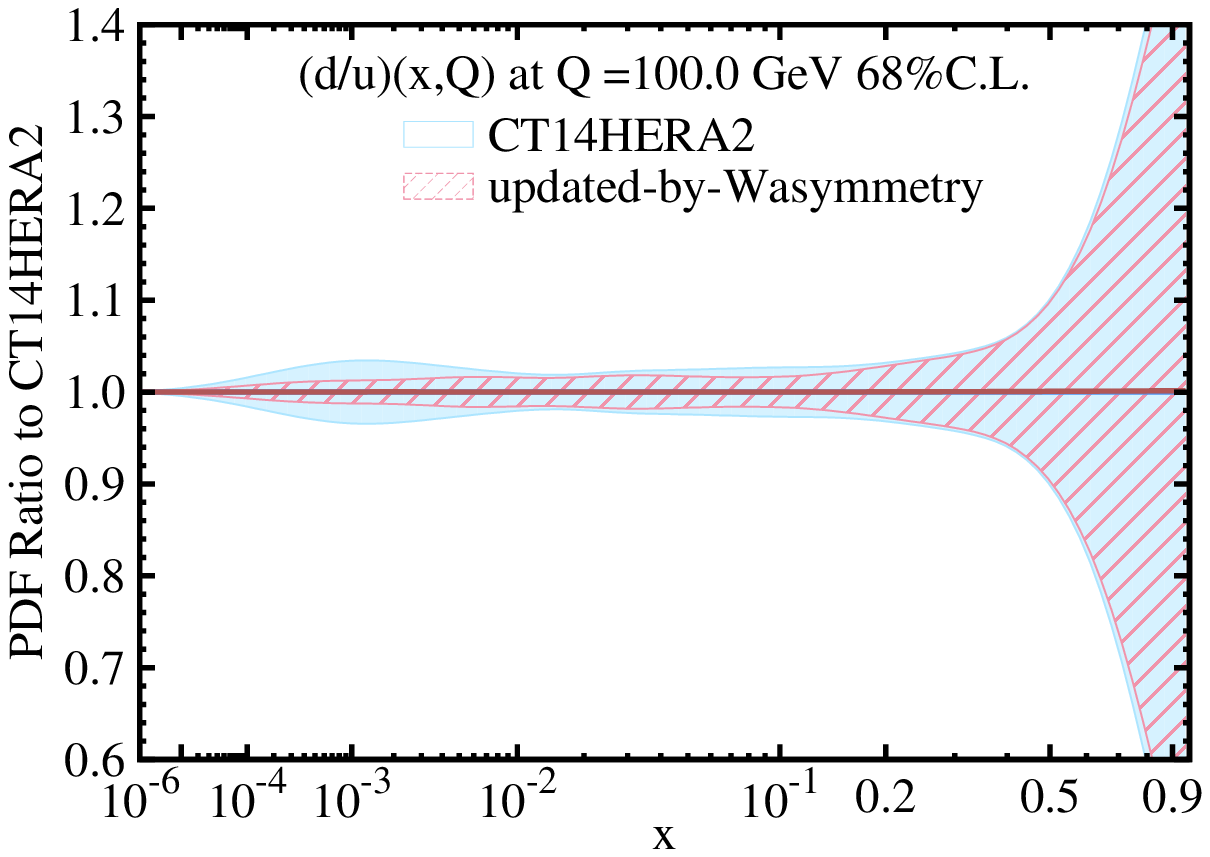}
\includegraphics[width=0.4\textwidth]{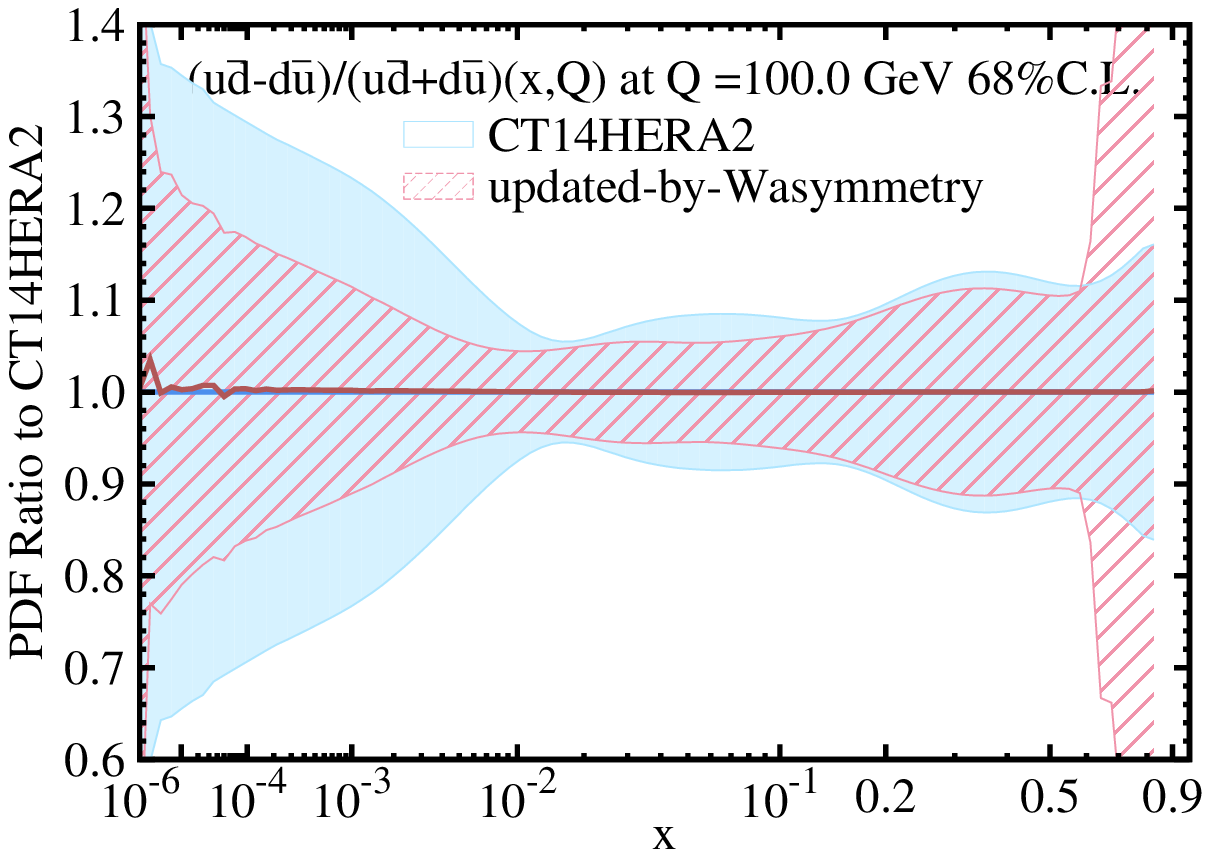}
\caption{\small 
Same as Fig.~\ref{fig:sidebandmassPDF}, but 
the lepton charge asymmetry  $A_\pm(\eta_\ell)$ data is used in the updating.
}
\label{fig:WupdatePDF}
\end{center}
\end{figure}

Fig.~\ref{fig:WupdatePDF} depicts the comparison of $d$, ${u_v}-{d_v}$, $d/u$ and ${(u{\bar d}-d{\bar u})/(u{\bar d}+d{\bar u})}$ PDFs, between the nominal CT14HERA2 and the updated PDFs with the inclusion of the 
$A_\pm(\eta_\ell)$  data.
It shows that the potential bias on the central values of the PDFs is negligible, while noticeable reduction of the PDF uncertainty can be clearly observed in some relevant $x$ ranges, depending on the parton flavor.   

\section{Updating PDFs using both sideband $A_{FB}$ and lepton charge asymmetry $A_\pm(\eta_\ell)$}
\label{sec:combined}

Since the Drell-Yan $A_{FB}$ and the lepton charge asymmetry $A_\pm(\eta_\ell)$ provide complementary information,
it is expected that the PDF-induced uncertainty in the determination of $\effstw$ can be further reduced if we use both 
the sideband $A_{FB}$ and the $A_\pm(\eta_\ell)$ data together to update the PDFs. 
Applying the same analysis to those two pseudo-data sets, as detailed in the previous sections, we obtain the results listed in  
Table~\ref{tab:sidebandAFB_Wasym_stw}, which should be directly compared to 
Table~\ref{tab:sidebandAFB_stw} for using the sideband $A_{FB}$ data and Table~\ref{tab:WasymmetryupdateResult} for using the 
$A_\pm(\eta_\ell)$ data alone, respectively.
We find that using both data sets to update the PDFs could further reduce the PDF-induced uncertainty on the determination of $\effstw$, 
which is determined using the $A_{FB}$ data in the $Z$-pole mass window, by about $28\%$ as compared to that using only the sideband 
$A_{FB}$ data.

\begin{table}[H]
\begin{center}
\begin{tabular}{|l|c|c|}
\hline \hline
Update PDF using & Potential bias & PDF uncertainty\\ 
sideband $A_{FB}$ and $A_\pm(\eta_\ell)$ & on $\effstw$ & on $\effstw$ \\
\hline
$CC$ events: & 0.00009 & 0.00032 \\
\hline
$CF$ events: & 0.00018 & 0.00021 \\
\hline \hline
\end{tabular}
\caption{\small 
 The bias and the PDF-induced uncertainty on $\stw$, after updating the PDFs with the sideband range of the  $A_{FB}$ pseudo-data (with $\stw=0.2324$) and the$A_\pm(\eta_\ell)$ pseudo-data samples, for the $CC$ and $CF$ event samples, respectively.
The PDF uncertainties are given at the 68\% C.L.
}
\label{tab:sidebandAFB_Wasym_stw}
\end{center}
\end{table}

\section{An application of ePump-optimization}

To further discuss the improvement in PDF-induced uncertainties, we apply the ePump-optimization method of the \eP code in this section, in the combined analysis of the $A_{FB}$ and $A_\pm(\eta_\ell)$ data, 
(1) to demonstrate their complimentary roles in reducing the PDF uncertainty in the PDF-updating procedure; and 
(2) to investigate the optimal choice of bin size for studying the PDF-induced uncertainty in experimental observables related to those two individual data.

\subsection{Complimentary roles in reducing the PDF uncertainty}
\label{sec:roles}

The ePump-optimization (or PDF-rediagonalization) method is based on ideas similar to that used in the data set diagonalization
method developed by Pumplin~\cite{Pumplin:2009nm}.

For a set of new data points, the application constructs an equivalent set of eigenvector, which are orthogonal to each other 
in the PDF fitting parameter space, by re-diagonalizing the original Hessian error PDFs with respect to the given data. 
The total uncertainty calculated by the new eigenvectors is exactly identical to 
that calculated with the original error PDFs in the linear approximation assumed by the Hessian analysis. But, in addition, 
the new error PDF pairs are ordered by the magnitudes of their re-calculated eigenvalues, the sum of which should be identical 
to the total number of the given data points, as noted in Ref.~\cite{ePumpPaper}. 
That is to say the new eigenvectors can be considered as projecting the original error PDFs to the given data set, and be optimized 
or re-ordered so that it is easy to choose a reduced set that covers the PDF uncertainty for the input data set to any desired 
accuracy~\cite{ePumpPaper}.

As an example, after applying the ePump-optimization method to the CT14HERA2 PDFs for the sideband $A_{FB}$ and $A_{\pm}(\eta_\ell)$ data
sets, which contain 50 data points in total ({\it i.e.,} 25 bins in each case), we find that the top three new eigenvector pairs predominantly have eigenvalues 
of 25.2, 18.1 and 5.5, respectively, while the eigenvalues of remaining ones decrease rapidly after that.
The combination of these top 3 optimized error PDFs contributes up to $97.6\%$ in the total PDF variance of the 50 given data points. 
This ePump-optimization allows us  to conveniently use these 3 leading new eigenvectors, in contrast to applying the full 56 error sets of 
the CT14HERA2, to study the PDF-induced uncertainty on $A_{FB}$ and $A_\pm(\eta_\ell)$ or any other observable that is directly 
related to them.

\begin{figure}[h]
\begin{center}
\includegraphics[width=0.4\textwidth]{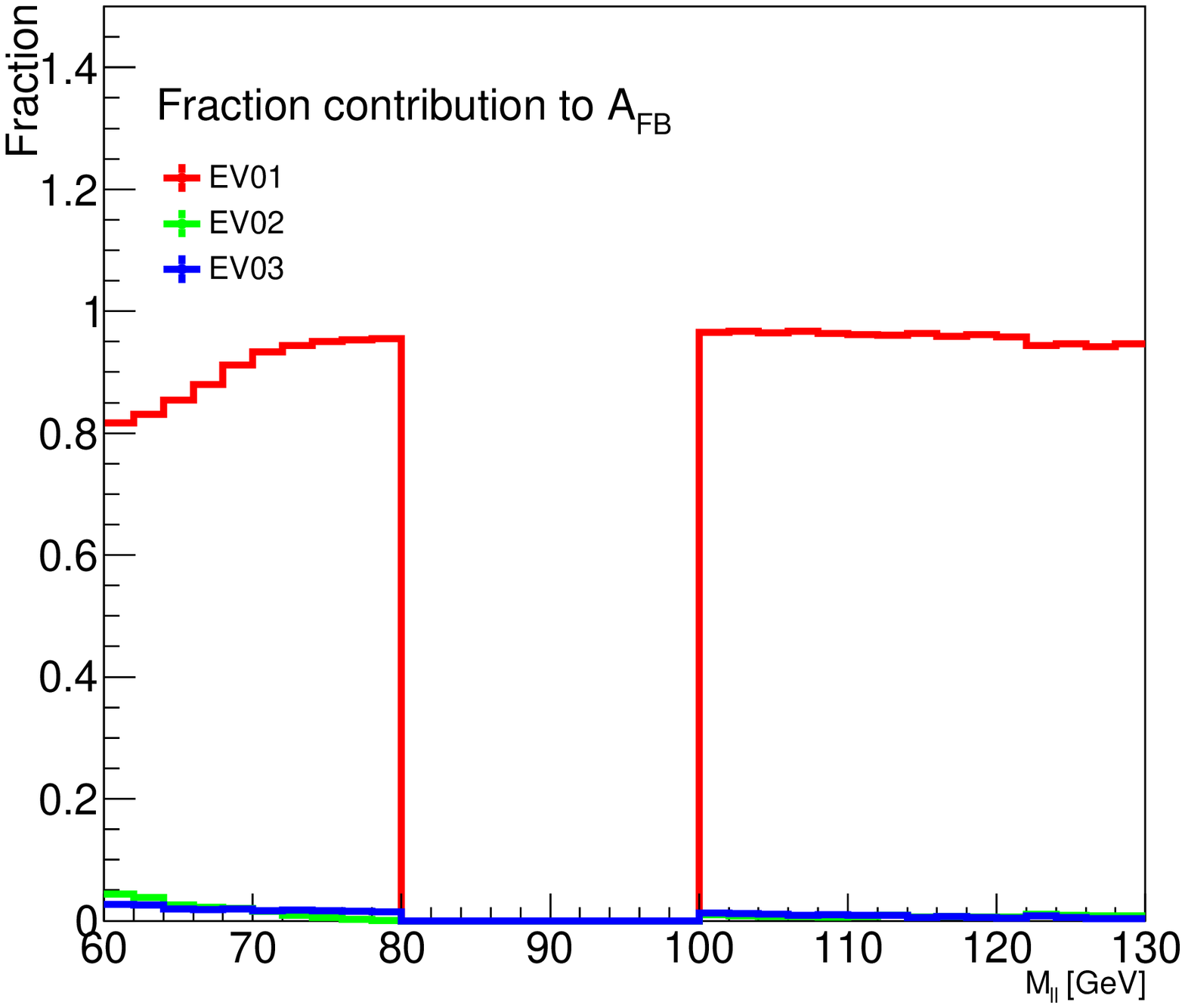}
\includegraphics[width=0.4\textwidth]{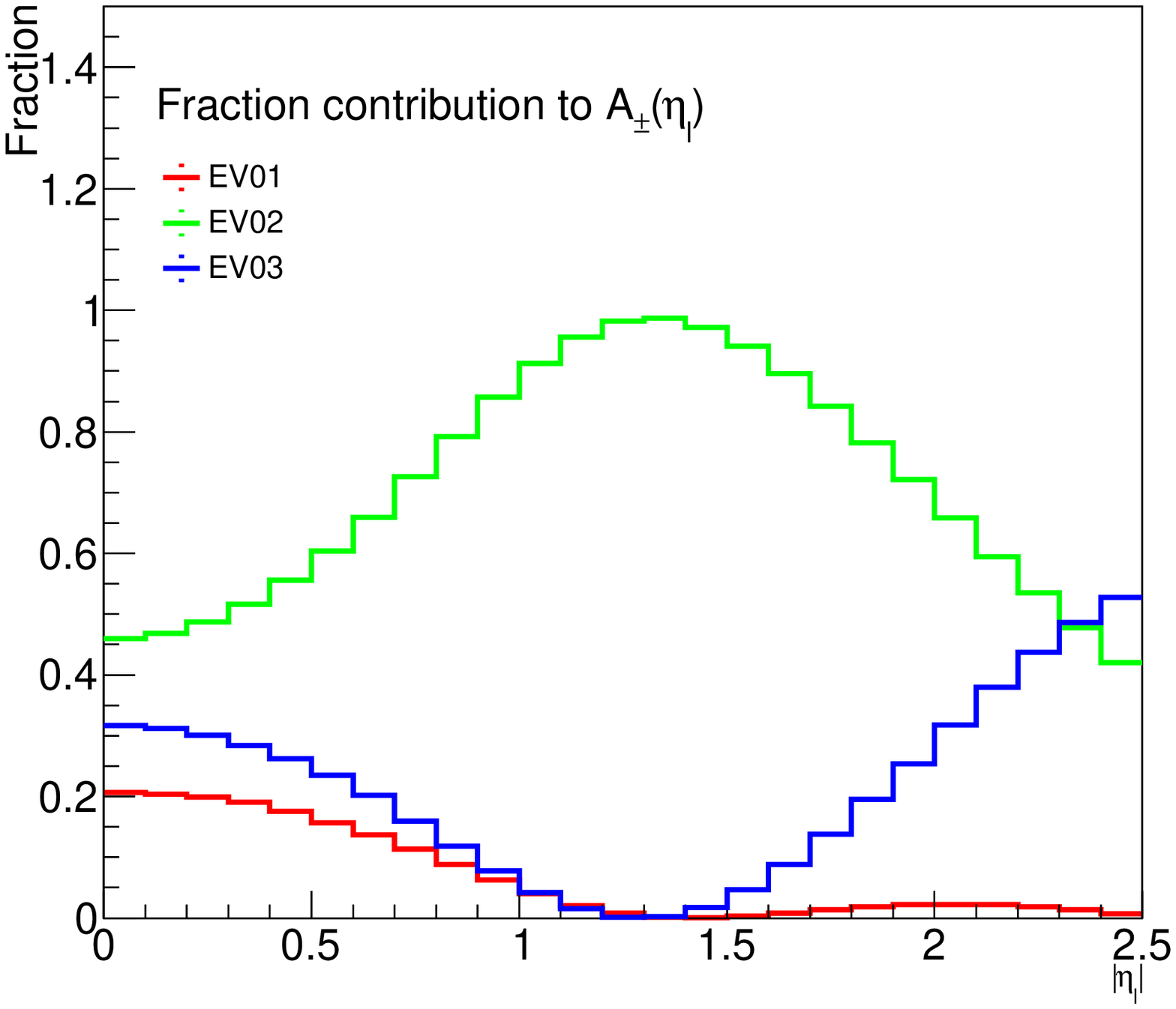}
\caption{\small 
Fractional contribution of the top three leading optimized eigenvectors (EV01, EV02 and EV03) to the variance of the observables
$A_{FB}$ and $A_\pm(\eta_\ell)$, normalized to each bin respectively, in the combined ePump-optimization analysis.}
\label{fig:fr_ev}
\end{center}
\end{figure}

Relative contributions of the top 3 leading optimized eigenvectors to the PDF uncertainties of the sideband $A_{FB}$ and $A_{\pm}(\eta_\ell)$,
normalized to each bin for illustration, are shown in Fig.~\ref{fig:fr_ev}, respectively. One can see directly that the first eigenvector (labeled as EV01) 
gives by far the largest contribution to the PDF uncertainties of the sideband $A_{FB}$, but very small fraction of the uncertainties 
of the $A_{\pm}(\eta_\ell)$, particularly for $|\eta_\ell|>1$; 
while the second and the third eigenvectors (labeled as EV02 and EV03) contribute a large or appreciable amount of 
the uncertainties on $A_{\pm}(\eta_\ell)$, but a much smaller fraction on $A_{FB}$. This suggests that when optimizing PDFs using both the $A_{FB}$ and $A_{\pm}(\eta_\ell)$ samples, 
these two data sets play complimentary roles in reducing the PDF uncertainties, {\it i.e.,} the re-diagonalization of the first pair of eigenvector 
is dominated by the information from the $A_{FB}$ and the second pair has more information from the $A_{\pm}(\eta_\ell)$.

\begin{figure}[hbt]
\begin{center}
\includegraphics[width=0.4\textwidth]{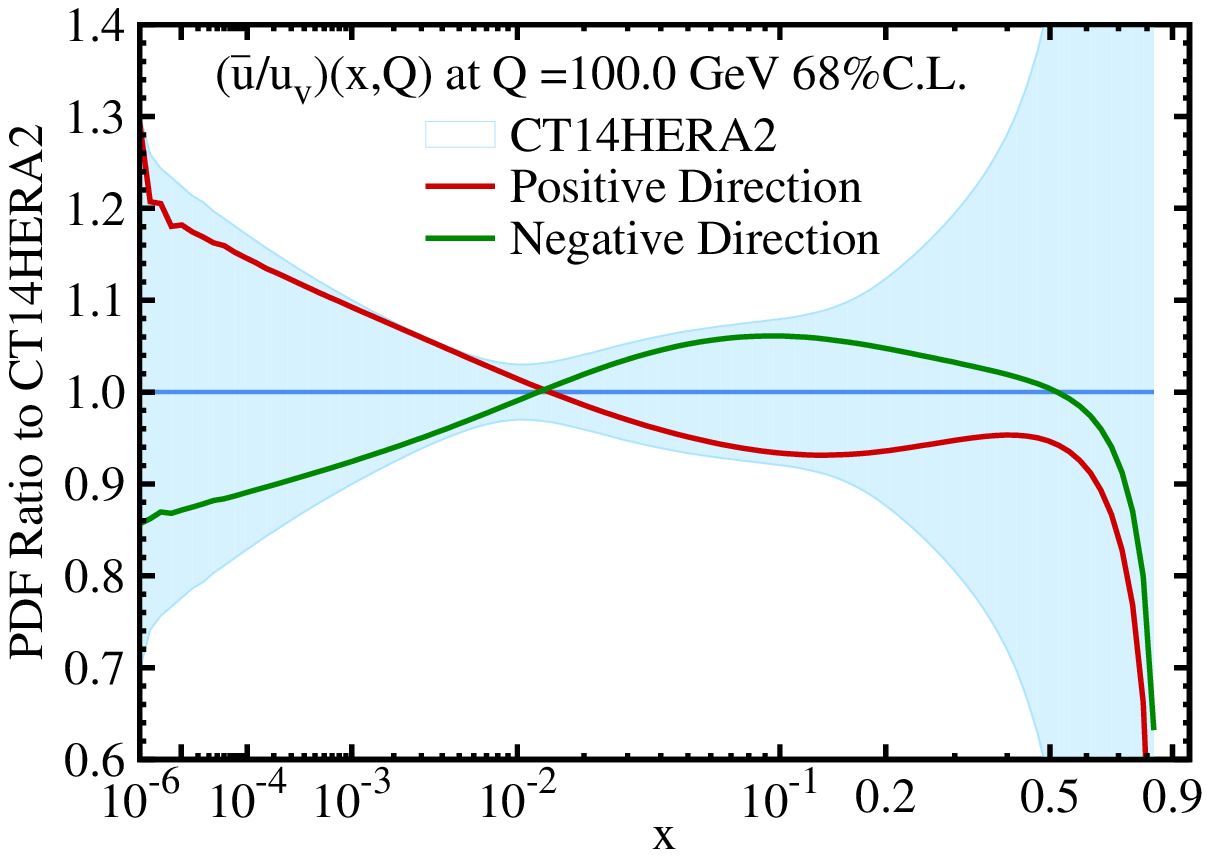}
\includegraphics[width=0.4\textwidth]{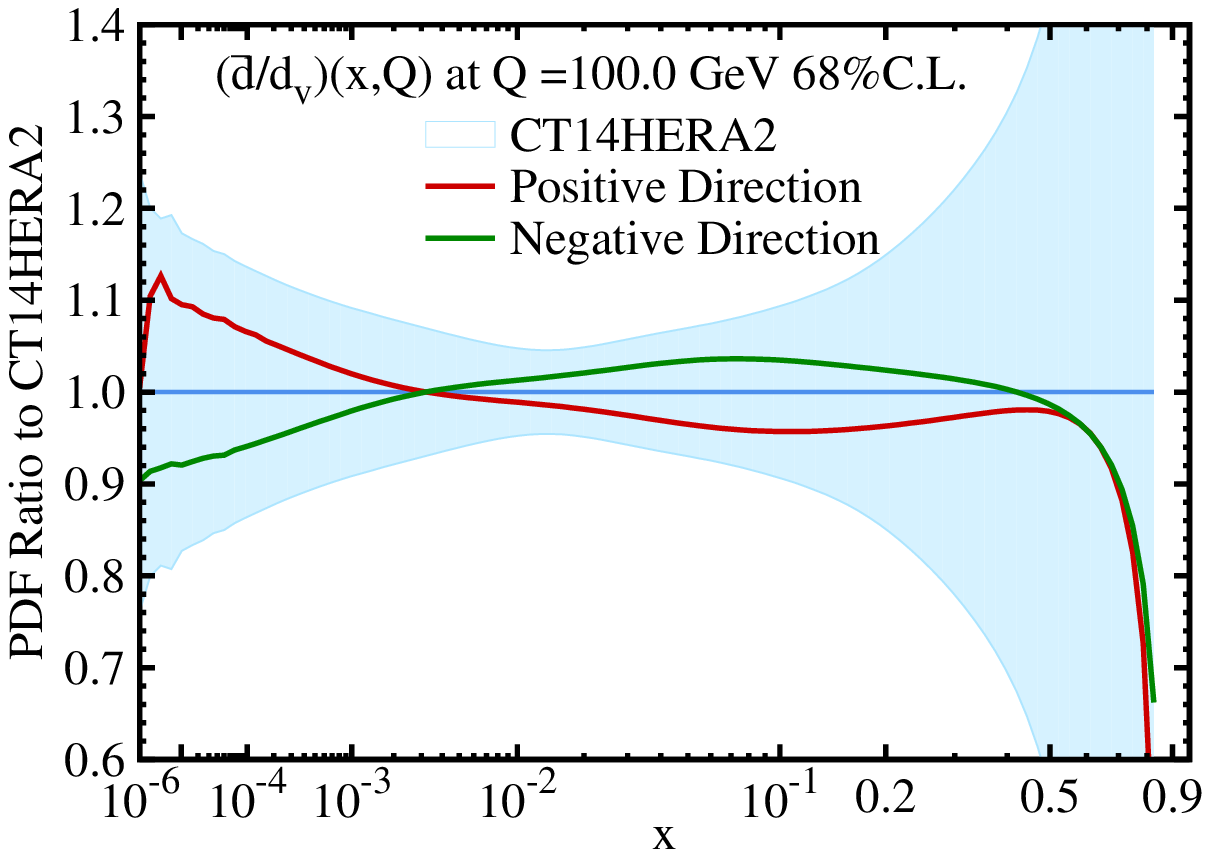}
\includegraphics[width=0.4\textwidth]{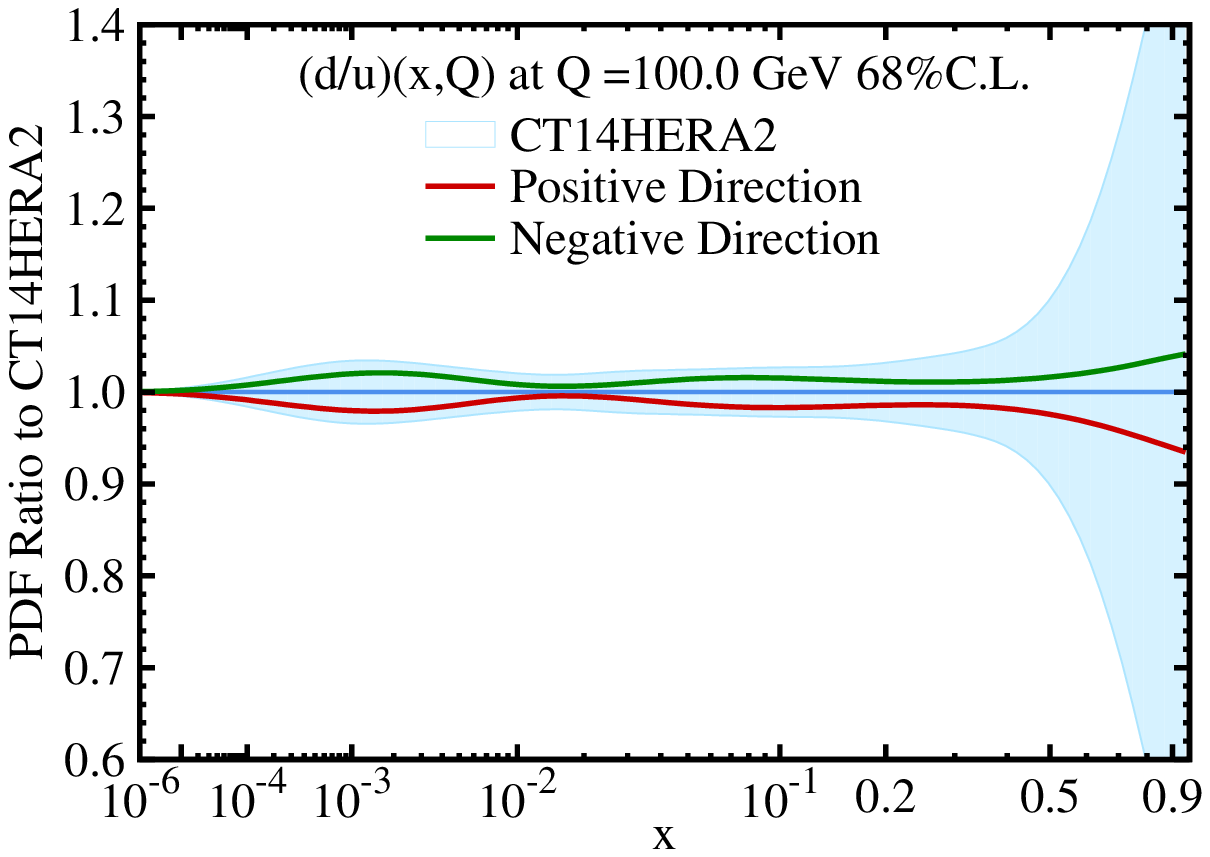}
\includegraphics[width=0.4\textwidth]{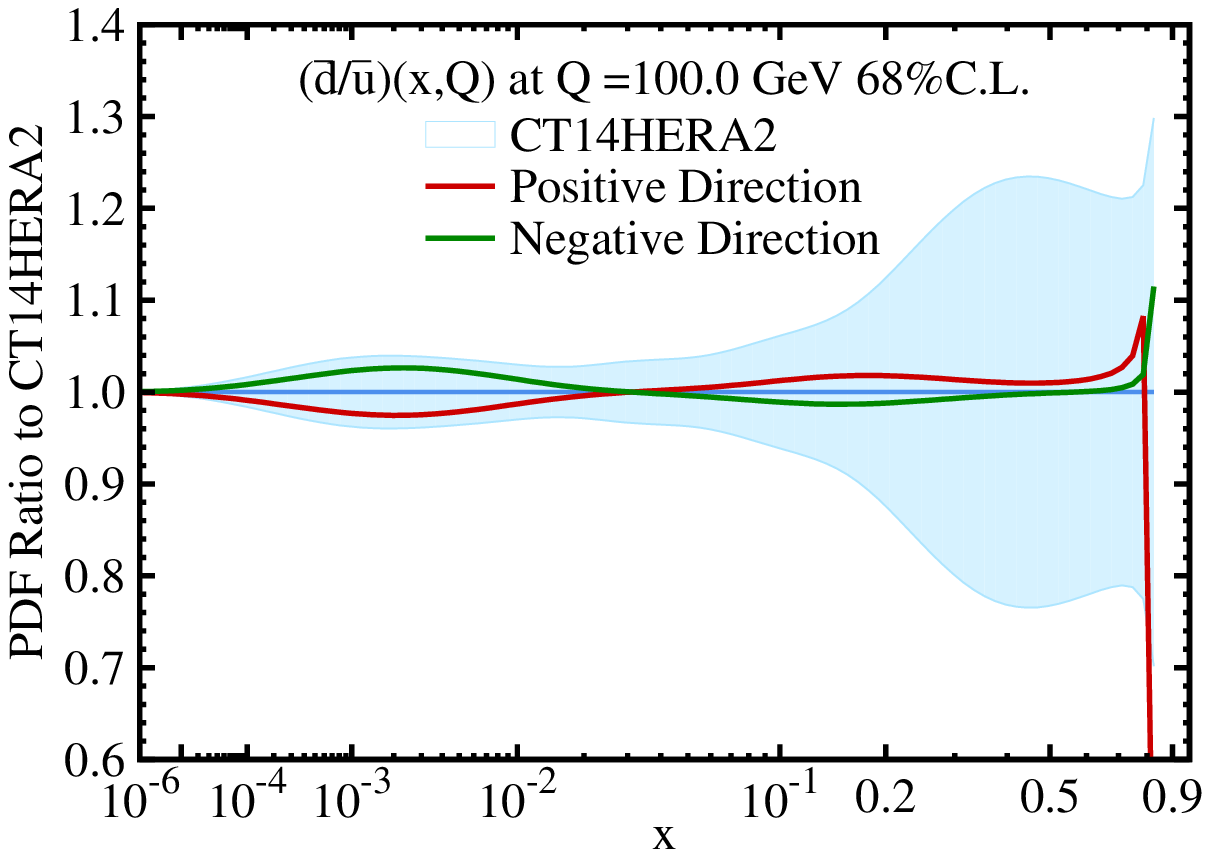}
\caption{\small Ratios of the 
first pair of eigenvector PDFs and the original CT14HERA2 error PDFs, at $Q=100$ GeV, to the CT14HERA2 central value of the ${\bar u}/{u_v}$, ${\bar d}/{d_v}$, $d/u$ and $\bar d/\bar u$ PDFs.
Those eigenvector PDFs were obtained after applying the ePump-optimization to the original CT14HERA2 PDFs 
in the combined analysis of the Drell-Yan sideband $A_{FB}$ and the lepton charge asymmetry  $A_\pm(\eta_\ell)$ data. 
}
\label{fig:eigenvector_first}
\end{center}
\end{figure}

\begin{figure}[hbt]
\begin{center}
\includegraphics[width=0.4\textwidth]{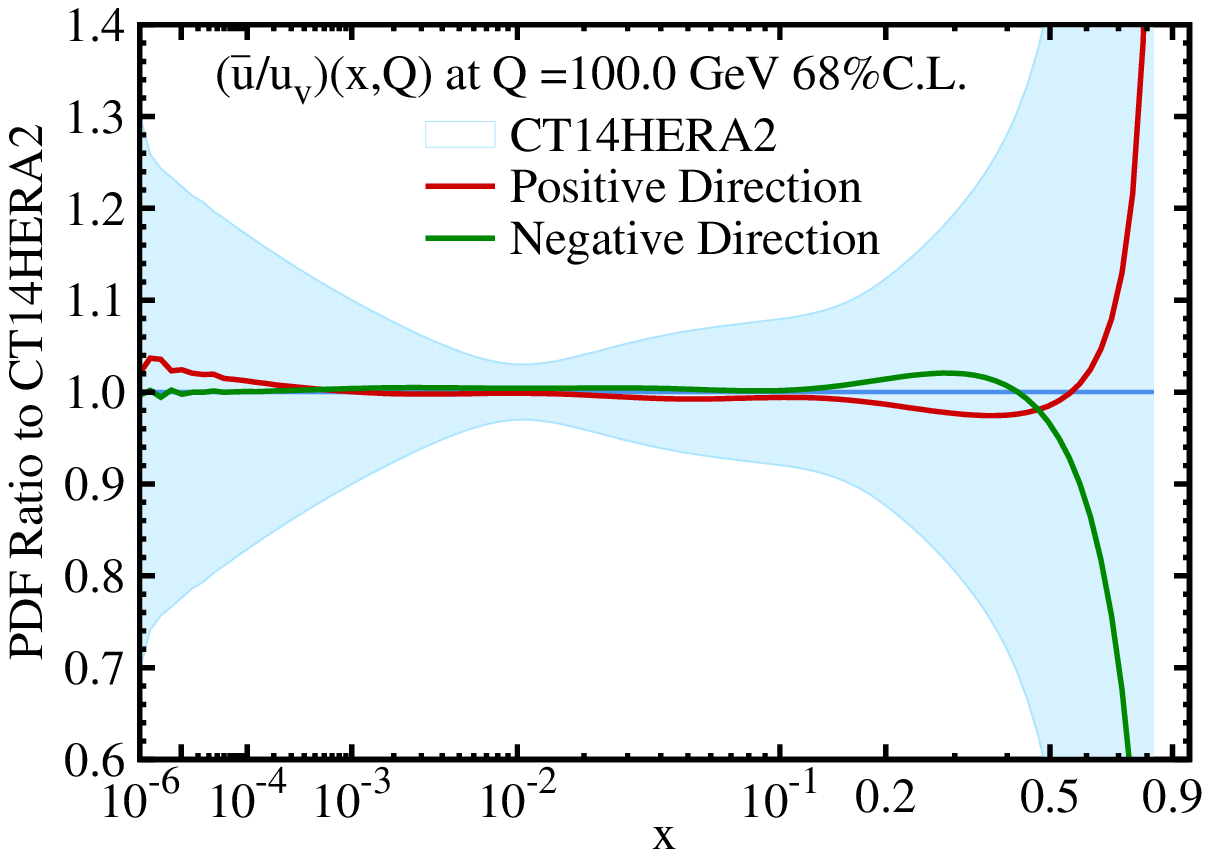}
\includegraphics[width=0.4\textwidth]{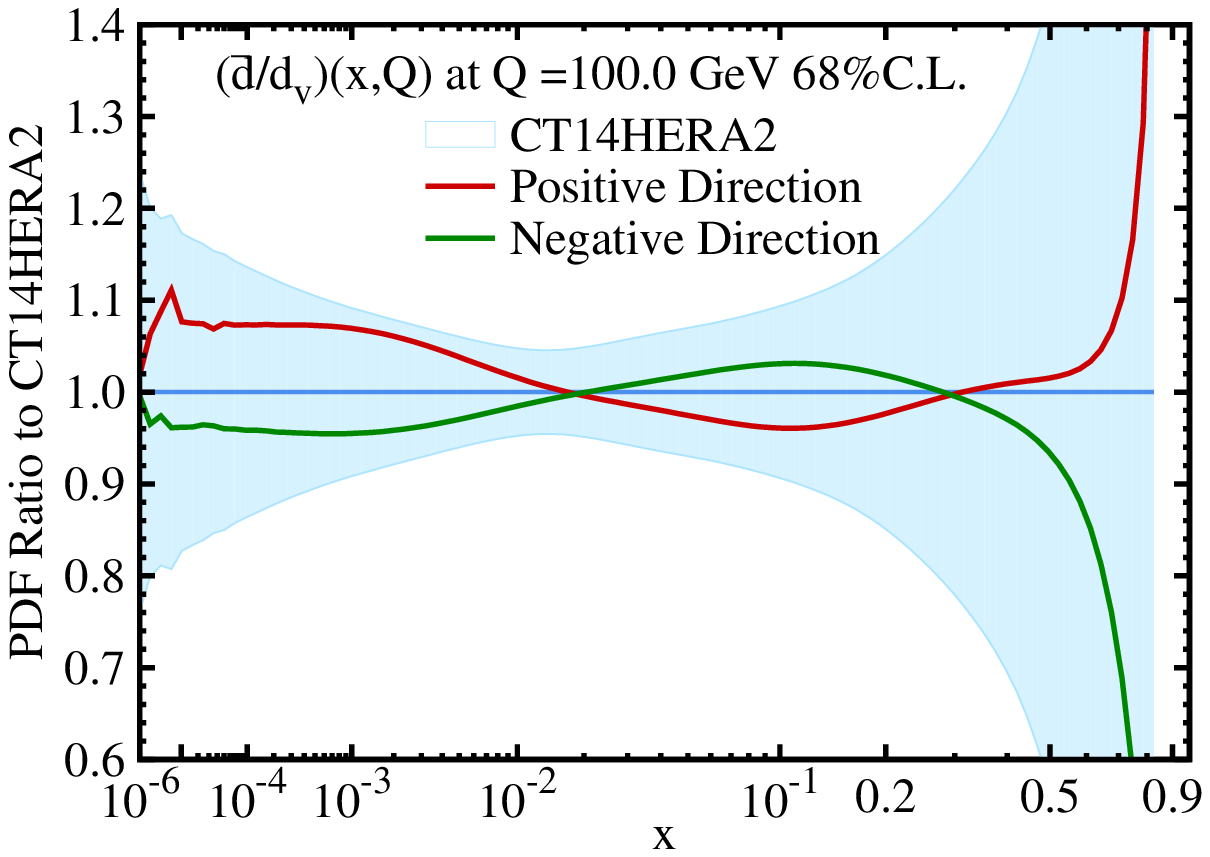}
\includegraphics[width=0.4\textwidth]{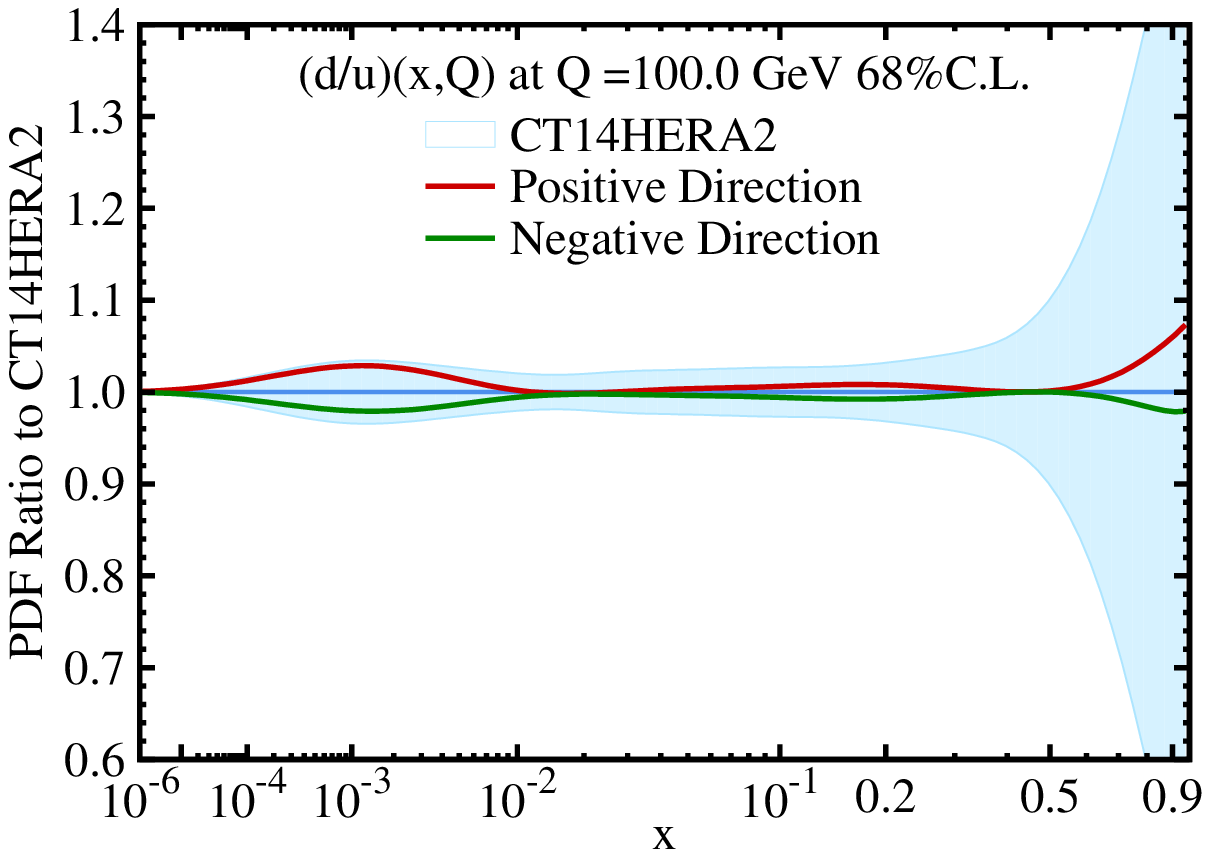}
\includegraphics[width=0.4\textwidth]{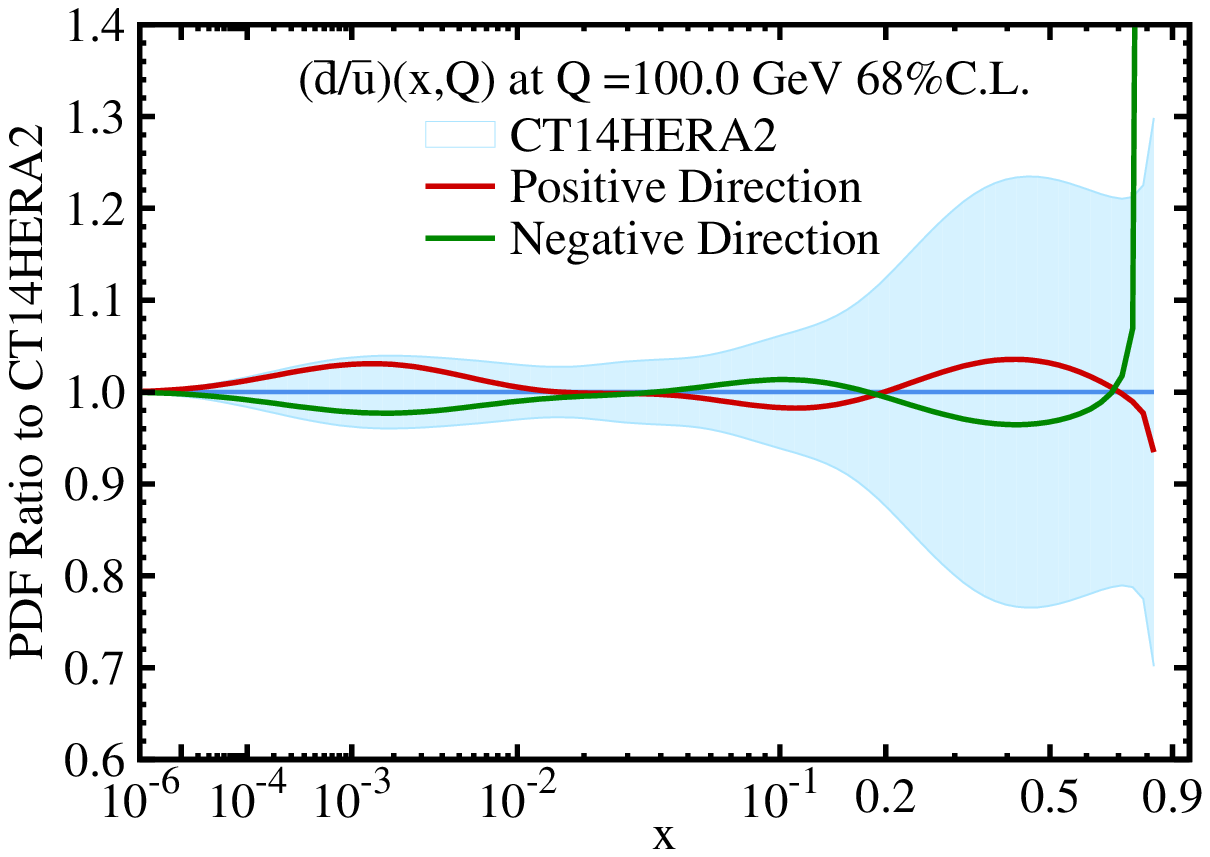}
\caption{\small 
Same as  Fig.~\ref{fig:eigenvector_first}, but for the second pair of eigenvector PDFs.
}
\label{fig:eigenvector_second}
\end{center}
\end{figure}

The sensitivities provided by the top two pairs of eigenvector PDFs to the different flavor and $x$-range, probed by the sideband $A_{FB}$ and the lepton charge asymmetry $A_{\pm}(\eta_\ell)$ together, are depicted in Fig.~\ref{fig:eigenvector_first} and Fig.~\ref{fig:eigenvector_second}, respectively. 
It could be verified that these two leading pairs of error PDFs, optimized by using both of those two data sets, resemble the respective first pair 
of eigenvector PDFs after applying the ePump-optimization procedure to the sideband $A_{FB}$ and the $A_{\pm}(\eta_\ell)$ alone.
This information can be understood from the following physical argument: $A_{FB}$ is dependent on PDFs predominantly because the dilution effect could lead to an incorrect assignment of the $z$ direction of the Collins-Soper definition. 
At the LHC, the leading order dilution probability that forward and backward is misjudged depends only on the relative size of PDF ratios $u/\bar{u}$ and $d/\bar{d}$, meaning that it is more sensitive to the quark-antiquark comparison.
For $A_{\pm}(\eta_\ell)$,this asymmetry comes from the difference between the $u\bar{d}$ cross section and the $d\bar{u}$ cross section, meaning that it is more sensitive to the flavor difference.
As shown in Fig.~\ref{fig:eigenvector_first}, the first eigenvector pair, which gives the largest PDF contribution to the $A_{FB}$ uncertainty,
dominates the $u/\bar{u}$ uncertainty in the $x$ region of a $Z$ boson process.
The $d/\bar{d}$ uncertainty is not as dominated by the first eigenvector, because the $Z$-quark couplings of neutral vector current, which
govern the magnitude of $A_{FB}$ at parton level, are proportional to the electric charges of different quark types, 
so that the sensitivity of $A_{FB}$ to $d/\bar{d}$ parton distribution is suppressed. 
Since the observed $A_{FB}$ is a combination of $u\bar{u}$ and $d\bar{d}$ processes, it can provide some information on the difference 
between $u$ and $d$ quark PDFs, but it is not as sensitive to this as it is to the $u/\bar{u}$ and $d/\bar{d}$ ratios.
In Fig.~\ref{fig:eigenvector_second}, the second eigenvector pair, which gives the largest PDF contribution to the $A_{\pm}(\eta_\ell)$ uncertainty, dominates the $d/u$ and $\bar{d}/\bar{u}$ uncertainties in the $x$ region of the single $W$ boson process. However, it has almost no sensitivity to 
the $\bar{u}/u_{v}$ uncertainty in a very large $x$-range.

\subsection{Optimal choice of bin size}

In previous sections, a bin size of 2 GeV on mass is used for measuring the $A_{FB}$ distribution, and a bin size of 0.1 on  $\eta_\ell$ is used for $A_\pm(\eta_\ell)$.
In principle, using a large bin size would smear some fine structures of the $A_{FB}$ and $A_\pm(\eta_\ell)$ distributions, and make those observables less sensitive to variations of PDFs. Hence, it is desirable to determine the maximally allowed bin size for a given observable 
without losing the sensitivity.  
Due to the difficulty in the experimental unfolding procedure to remove detector effects, such as bin-to-bin migration effects 
and determination of efficiency and acceptance, 
it may not be always practical to measure the $A_{FB}$ and $A_\pm(\eta_\ell)$ distributions in such a fine bin configuration. 
In this section, we discuss how to apply the ePump-optimization procedure to obtain the optimal choice of bin size for 
$A_{FB}$ and  $A_\pm(\eta_\ell)$ distributions.

\begin{figure}[hbt]
\begin{center}
\includegraphics[width=0.4\textwidth]{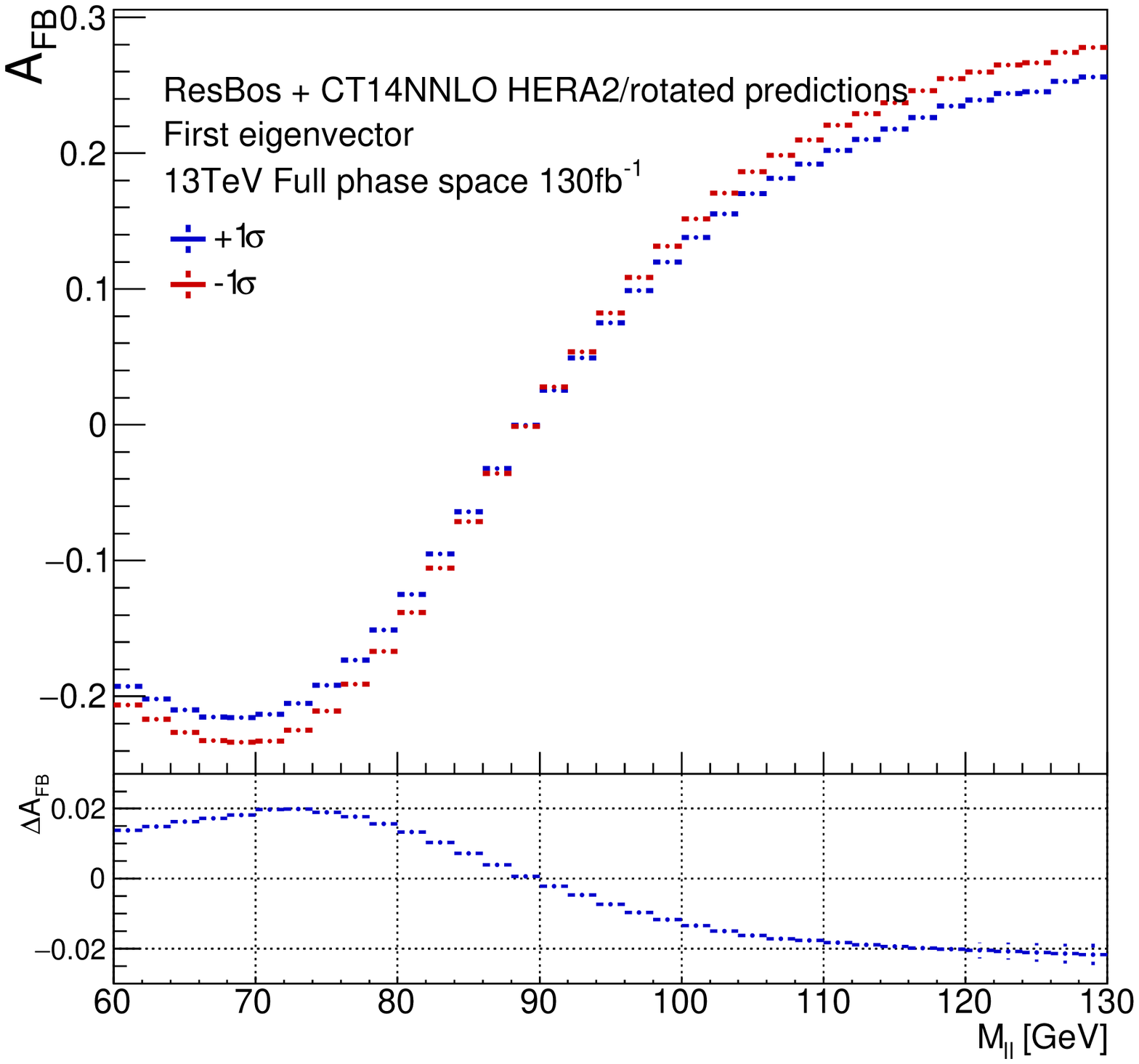}
\includegraphics[width=0.4\textwidth]{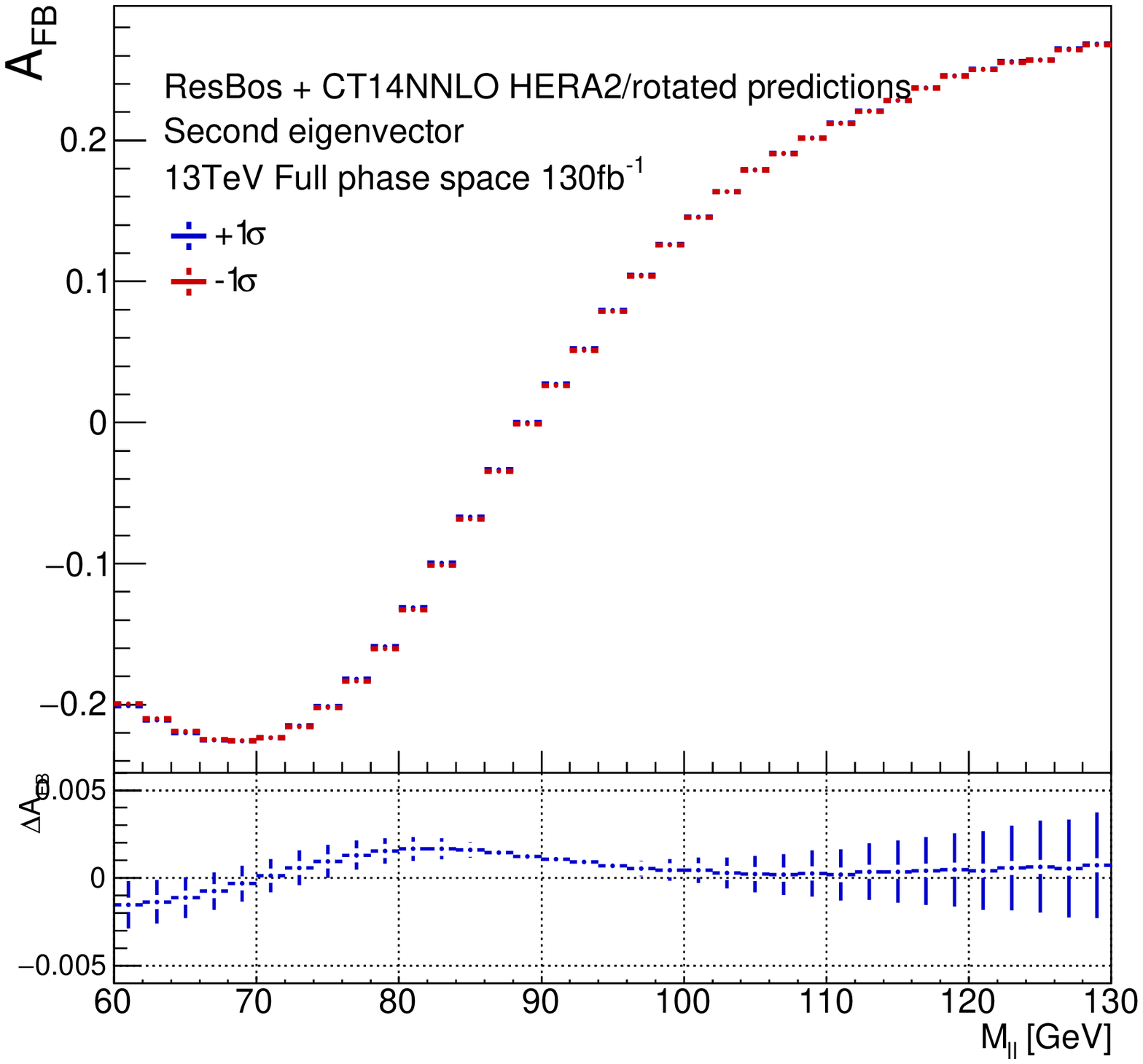}
\caption{\small $A_{FB}$ distribution predicted by the first and second eigenvector PDF sets, after applying the ePump-optimization method to optimize the CT14HERA2 PDFs for the  $A_{FB}$ data, as described in the text. 	
Predictions from the positive and negative shifted error sets of each eigenvector PDF set are compared, and 
$\Delta A_{FB}$ is their difference.
}
\label{fig:PDFunc_AFB}
\end{center}
\end{figure}

\begin{figure}[hbt]
\begin{center}
\includegraphics[width=0.4\textwidth]{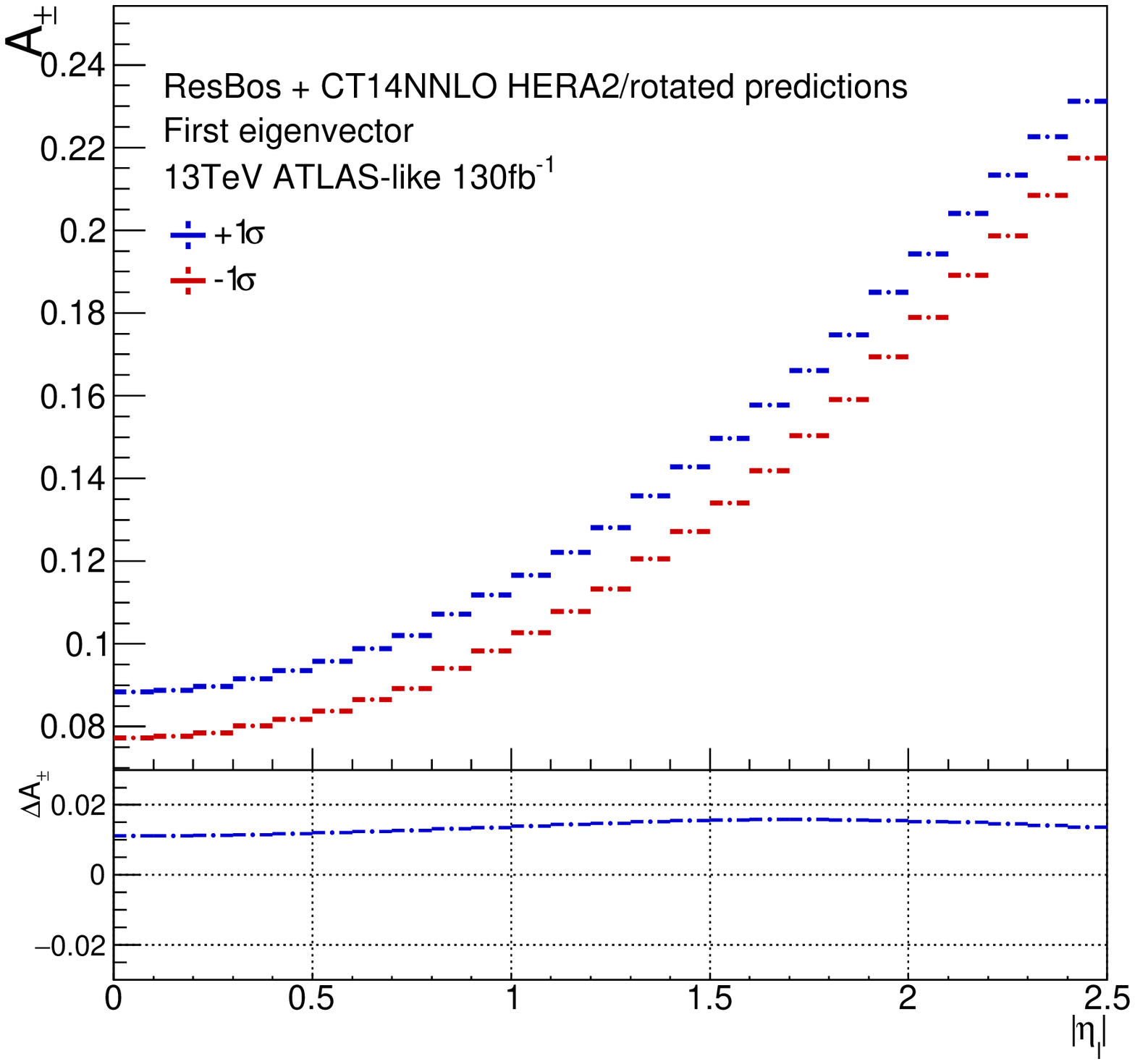}
\includegraphics[width=0.4\textwidth]{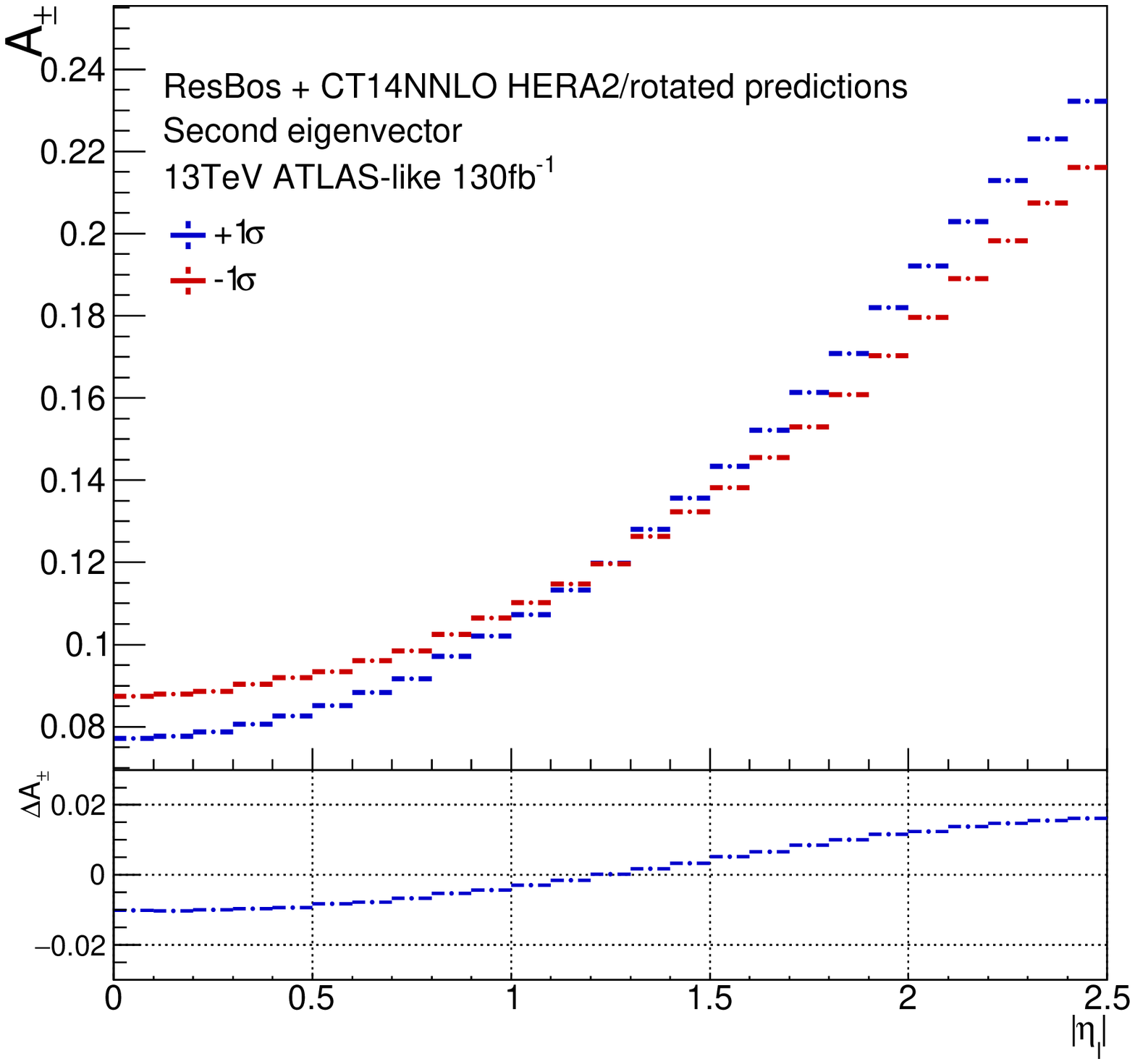}
\caption{\small 
Similar to Fig.~\ref{fig:PDFunc_AFB}, but for 	
$A_\pm(\eta_\ell)$distribution.
}
\label{fig:PDFunc_Wasy}
\end{center}
\end{figure}

From Sec.~\ref{sec:roles}, we learned that the PDF-induced error on $A_{FB}$ and $A_\pm(\eta_\ell)$ can be represented 
by the leading eigenvectors after ePump-optimization. In Fig.~\ref{fig:PDFunc_AFB}, we show the $A_{FB}$ distributions, 
predicted by the first two eigenvector PDF sets, after PDF-rediagonization.For each eigenvector, positive and negative 
shifted PDF error sets are compared. Similarly, for the $A_\pm(\eta_\ell)$ distribution, comparisons are shown in Fig.~\ref{fig:PDFunc_Wasy}.

When the PDFs are varied according to the first pair of eigenvector sets, the most significant change in the shape of $A_{FB}$ distribution occurs 
as an oppositely shifted effect in high mass and low mass regions around the $Z$-pole, cf. the left-hand plot of Fig.~\ref{fig:PDFunc_AFB}. 
Moreover, the shape of $\Delta$ is almost flat either below or above the $Z$-pole mass window, 
where $\Delta$ is the difference between the two values of $A_{FB}$ predicted by the positive and negative shifted error sets. 
As a result, using a larger bin size on mass will not lose much information on how PDFs affect $A_{FB}$ distribution.  
On the other hand, as shown in the right-hand plot of Fig.~\ref{fig:PDFunc_Wasy}, when the PDFs are varied according to the second pair of eigenvector sets, 
the change of $\Delta$ in the shape of $A_\pm(\eta_\ell)$ distribution is almost a linear-type. Hence, as long as the bin size on 
lepton rapidity still reflects the linear shape, the sensitivity of $A_\pm(\eta_\ell)$ to PDF variations should not be dramatically reduced.

To quantitatively study the sensitivity loss originated from using a larger bin size, we compare to another analysis done by using 
a bin size of 5 GeV on mass for the $A_{FB}$ distribution, and a bin size of 0.25 on lepton $\eta$ for the $A_\pm(\eta_\ell)$ distribution, 
for the same data samples as used in Section ~\ref{sec:combined}.
We find that numerical calculations by using these wide bins give exactly the same results as that presented in the previous tables, 
which implies that the reduction of PDF 
uncertainty would not be compromised by using larger bin size, as proposed above.
This leads to a very useful conclusion: aiming for the $\effstw$ measurement, both $A_{FB}$ and $A_\pm(\eta_\ell)$ distributions can be measured in 
a large bin size to reduce systematic uncertainties without losing much sensitivity in constraining the PDFs. This conclusion should hold for both a quick 
PDF-updating and a full PDF global fitting. This conclusion is important, 
because as more data accumulates at the LHC, systematic uncertainties will soon 
be larger than the statistical uncertainty for many precision measurements. Therefore reducing systematics should have 
higher priority.
 
\iffalse
\begin{table}[hbt]
\tiny{
\begin{center}
\begin{tabular}{l|c|c|c|}
\hline \hline
    Updating & Observable & PDF uncertainty & PDF uncertainty \\
     & & before updating & after updating \\
\hline
  $A_{FB}$ 2 GeV / bin & $CC$ average $A_{FB}$ & 0.00062  & 0.00036 \\
  & $CF$ average $A_{FB}$ & 0.00194 & 0.00119 \\
\hline
  $A_{FB}$ 5 GeV / bin &  $CC$ average $A_{FB}$ & 0.00062 & 0.00036 \\
   & $CF$ average $A_{FB}$ & 0.00194 & 0.00119 \\
\hline
  $A_\pm(\eta_\ell)$ 0.1 / bin & $CC$ average $A_{FB}$ & 0.00062 &  0.00051 \\
   & $CF$ average $A_{FB}$ & 0.00194 & 0.00169 \\ 
\hline
  $A_\pm(\eta_\ell)$ 0.25 / bin & $CC$ average $A_{FB}$ & 0.00062 & 0.00051 \\
  & $CF$ average $A_{FB}$ & 0.00194 & 0.00169 \\ 
\hline \hline
\end{tabular}
\caption{\scriptsize PDF updating using different bin size.}
\label{tab:binsize}
\end{center}
}
\end{table} 
\fi

\section{Summary}

We have presented a study on how to correctly reduce the PDF-induced uncertainty in the determination of the effective weak mixing angle $\effstw$, 
obtained from analyzing the measurement of the Drell-Yan forward-backward asymmetry $A_{FB}$ at the LHC.
According to previous studies, the PDF-induced uncertainty can be reduced by the PDF updating procedure using $A_{FB}$. 
However, when $A_{FB}$ is used for both PDF updating and $\effstw$ extraction, the correlation between these two important tasks 
will cause bias on both the updated PDFs and the extracted value of $\effstw$. Considering the deviation between the previous precise measurements on $\effstw$, 
such bias could be at the same level as the PDF-induced uncertainty on $\effstw$. 
In this paper we have shown how this bias can be suppressed.
$A_{FB}$ is more sensitive to $\effstw$ around the $Z$ pole, while the PDFs affect $A_{FB}$ more significantly in the sideband regions such as e.g. $60 < M_{ll}<80$ GeV and $100 < M_{ll}<130$ GeV. 
Accordingly, we propose to use the sideband $A_{FB}$ to reduce the correlation between the $\effstw$ extraction and the PDF updating, 
so that the bias on the $\effstw$ determination can be suppressed, while not significantly losing sensitivity in the PDF updating.

We have applied the \eP program, based on the Hessian updating method, to update the CT14HERA2 PDFs to update the CT14HERA2 PDFs by including the full mass range $A_{FB}$ 
pseudo data as new input to a global PDF fitting.With this updated-PDF set, we analyzed the extraction of $\effstw$ in the $Z$-pole mass window and found a sizable bias 
in its value, with respect to its input value in the pseudo data. Furthermore, the central values of the updated $d$ and $u$ quark PDFs, obtained from this analysis, 
are different from that of the original CT14HERA2 PDFs. This is caused by the difference in the $\effstw$ values assumed in the pseudo data and the theory templates. 
To reduce this type of correlation, we proposed to use only the sideband $A_{FB}$ to update the existing PDFs. 
As expected, using only the sideband $A_{FB}$ data to update the PDFs reduces the bias on the extraction of $\effstw$ value as well as the central values of the updated PDFs. 
We also show that the asymmetry from $W$ boson decay, $A_\pm(\eta_\ell)$ can be used to further reduce the PDF uncertainty. 
It plays a complementary role to the sideband $A_{FB}$ data in reducing the PDF-induced uncertainty, with negligible bias on the determination of the weak mixing angle.

A study on the effect of choosing different bin sizes of the $A_{FB}$ and $A_\pm(\eta_\ell)$ distributions was also performed. 
It showed that using somewhat larger bin size will not sacrifice 
much of the sensitivity of those two observables in reducing the PDF uncertainty in the $\effstw$ measurement. When more data are accumulated at the LHC, the systematical 
uncertainties in the $A_{FB}$ and $A_\pm(\eta_\ell)$ measurements will begin to dominate. In that case, there is an advantage in choosing a larger bin size in order to reduce 
the systematical uncertainties in the experimental unfolding procedures. In this study, using a bin size of 5 GeV on mass for the $A_{FB}$ distribution, and a bin size of 0.25 
on lepton $\eta$ for the $A_\pm(\eta_\ell)$ distribution 
did not cause a noticeable reduction in the sensitivity of these two data sets to the measurement of $\effstw$.

In conclusion, we have investigated the correlation and potential bias in reducing the PDF-induced uncertainty in the determination of $\effstw$ from the forward and backward 
asymmetry $A_{FB}$ of the Drell-Yan processes at the LHC. Derived from quantitative computation of the Hessian-based \eP PDF updating program, it can be concluded that 
by excluding $Z$ pole region events in the PDF updating, the potential bias on the $\effstw$ extraction would not significantly enlarge the estimated 
total uncertainty, including the statistical and PDF-induced uncertainties at the LHC Run 2.
However, the bias is not negligible and thus still needs careful evaluation in the future precise $\effstw$ measurements at the high luminosity LHC. 
Moreover, although it is useful to quickly use \eP to estimate the impact of a new data set on the PDFs, we suggest to use the PDF updating method as only a preliminary way 
to reduce the PDF-induced uncertainty in the $\effstw$ measurements. A full PDF global fitting analysis is necessary for a complete determination of $\effstw$ with PDF correlations, 
in which new degrees of freedom in the non-perturbative parametrization of the PDFs can be explored. 
Furthermore, all experimental results, {\it i.e.,} $A_{FB}$ and $A_\pm(\eta_\ell)$ studied in this article, should be provided in a format such that theorists could replace the 
preliminary PDF updating method employed in the experimental analysis by a consistent global analysis.

\section{Acknowledgments}
\begin{acknowledgments}
This work was supported by the National Natural Science Foundation of China under Grant No. 11721505 and 11875245.
This work was also supported by the U.S. National Science Foundation under Grants No. PHY-1719914.
C.-P. Yuan is also grateful for the support from
the Wu-Ki Tung endowed chair in particle physics.
\end{acknowledgments}

\begin{appendix}
\section{Dilution effect on $A_{FB}$}
  
  Consider that if the directions of the initial state quarks and antiquarks in the DY events are known, one can 
  defined a Collin-Soper frame at hadron colliders without any dilution effect. Under this situation, the 
  differential cross section of the DY process can be written as:
  
\begin{eqnarray}
 \frac{d\sigma_q}{d\cos\theta^*_q} \sim (1+\cos^2\theta^*_q) + A^q_0 \times \frac{1}{2}(1-3\cos^2\theta^*_q) + A^q_4 \times \cos\theta^*_q
\end{eqnarray}
  where label $q$ is used to mark the no-dilution cross section at partonic level. In reality, the dilution effect can lead to an incorrect 
  assignment of the $z$ direction in the Collins-Soper frame, and resulting in:
  
\begin{eqnarray}
 \cos\theta^*_h = \cos(\pi - \theta^*_q) = -\cos\theta^*_q
\end{eqnarray}
  where label $h$ is used to mark the dilution case at hadronic level. With a dilution probability of $f$, we have:
  
\begin{eqnarray}
 \frac{d\sigma_h}{d\cos\theta^*_h} &=& f \times \frac{d\sigma_q}{d\cos\theta^*_q}\Big|_{\cos\theta^*_q = -\cos\theta^*_h} + 
 (1-f) \times \frac{d\sigma_q}{d\cos\theta^*_q}\Big|_{\cos\theta^*_q = \cos\theta^*_h} \nonumber \\
 &\sim&  (1+\cos^2\theta^*_h) + A^q_0 \times \frac{1}{2}(1-3\cos^2\theta^*_h) + (1-2f) \times A^q_4 \times \cos\theta^*_h
\end{eqnarray} 
  Accordingly,
\begin{eqnarray}
  A^h_4 &=& (1-2f)A^q_4 
\end{eqnarray}
  Since the $A_{FB}$ around $Z$-pole is proportional to the angular coefficient $A_4$, thus it turns out
\begin{eqnarray}
  A^h_{FB} &=& (1-2f) A^q_{FB}
\end{eqnarray}
  The sensitivity of constraining the PDFs (namely constraining $f$) via $A_{FB}$ depends on the $A_{FB}$ value itself.

\end{appendix}

\end{document}